\documentclass[aps,prb,showpacs,twocolumn,floatfix,nobalancelastpage,superscriptaddress,footinbib,longbibliography]{revtex4-1}

\usepackage[colorlinks=true,citecolor=blue,linkcolor=red,urlcolor=blue]{hyperref}
\usepackage{graphicx}
\usepackage{amsmath}
\usepackage{amssymb}
\usepackage{epstopdf}
\usepackage{color}

\usepackage[english]{babel}
\usepackage[applemac]{inputenc}

\usepackage{tikz}
\usetikzlibrary{decorations.pathmorphing}

\newcommand{\Arccos}{\textrm{Arccos}}

\renewcommand{\k}{{\vec k}}

\newcommand{\be}{\begin{equation}}
\newcommand{\ee}{\end{equation}}

\renewcommand{\r}{{\vec r}}

\newcommand{\p}{{\vec p}}

\newcommand{\tr}{{\rm tr}}
\newcommand{\ep}{\epsilon}

\newcommand{\gamDes}{\gamma}

\begin{document}

\title{Diffusion of Dirac fermions  across  a topological merging transition  in two dimensions}

\author{P. Adroguer}
\affiliation{Laboratoire de Physique, \'Ecole Normale Sup\'erieure de Lyon,
46 all\'ee d'Italie, 69007 Lyon, France}

\author{D. Carpentier}
\affiliation{Laboratoire de Physique, \'Ecole Normale Sup\'erieure de Lyon,
46 all\'ee d'Italie, 69007 Lyon, France}

\author{G. Montambaux}
\affiliation{Laboratoire de Physique des Solides, CNRS, Universit\'e Paris-Sud,
Universit\'e Paris-Saclay, 91405 Orsay Cedex, France}

\author{E. Orignac}
\affiliation{Laboratoire de Physique, \'Ecole Normale Sup\'erieure de Lyon,
46 all\'ee d'Italie, 69007 Lyon, France}

%
%
%\author{P. Adroguer$^{1,2}$, D. Carpentier$^{1}$, E. Orignac$^{1}$ and G. Montambaux$^3$}
%\affiliation{$^{1}$Laboratoire de Physique, \'Ecole Normale Sup\'erieure de Lyon,
%47 all\'ee d'Italie, 69007 Lyon, France,}
%\affiliation{$^{2}$Institute of Theoretical Physics and Astrophysics,
%University of W\"urzburg, D-97074 W\"urzburg, Germany,}
%\affiliation{$^{3}$Laboratoire de Physique des Solides, CNRS UMR 8502, Univ. Paris-Sud, F-91405 Orsay Cedex, France}
\date{\today}

\begin{abstract}
{
A continuous deformation of a Hamiltonian possessing at low energy two
Dirac points of opposite chiralities can lead to a gap opening by
merging of the two Dirac points. In two dimensions, the critical
Hamiltonian possesses a semi-Dirac spectrum: linear in one
direction but quadratic in the other.
We study the transport properties across such a transition, from a
Dirac semi-metal through a semi-Dirac phase towards a gapped
phase. Using both a Boltzmann approach and a diagrammatic Kubo
approach, we describe the conductivity tensor within the diffusive
regime. In particular, we show that both the anisotropy of the  Fermi
surface and the Dirac nature of the eigenstates combine to give rise
to anisotropic transport times, manifesting themselves through an
unusual matrix self-energy.
}
\end{abstract}

% We may need a few words on conductivity

\maketitle

 \section{Introduction}

The discovery of graphene has triggered a lot of work on the exotic transport properties of Dirac-like particles in solids \cite{castroneto2009}.
Indeed, the graphene electronic spectrum is made of two sub-bands
which touch at two inequivalent points in   reciprocal space.
Near the touching points, named Dirac points, the spectrum has a
linear shape and the electron dynamics is well described by a 2D
Dirac equation for massless particles.
Due to the structure of the honeycomb lattice, the wave functions have two components corresponding to the two inequivalent sites of the lattice, and the Hamiltonian is a $2 \times 2$ matrix.
To describe the low energy properties, the original Hamiltonian is replaced by two copies of a 2D Dirac equation
\be
H= \pm c \  \p  \, . \, {\vec \sigma}
 \ee
 where the velocity $c \simeq 10^5~$m.s$^{-1}$.
 This linearization is possible because the energy of the saddle point separating the two Dirac cones (valleys) is very large ($\simeq 3$ eV) compared to the Fermi energy and temperature scales.
Other realizations of Dirac-like physics in two dimensions have been proposed in the
organic conductor  (BEDT-TTF)$_2$I$_3$ under
pressure\cite{kobayashi:2007,goerbig:2008,suzumura:20011,choji:2011}, and  has been observed in
artificially assembled nanostructures\cite{Polini:13,bittner:2010,bellec:2013,rechtsman:2013,jacqmin:2014} and ultracold
atoms\cite{tarruell:2012,lim:2012}.
 Besides these two dimensional realizations, the existence and properties of semi-metallic phases in three dimensions have recently been studied
\cite{Vafek:2014,Hosur:2013}.

\begin{figure} [!h]
\centering
\includegraphics[width=8cm]{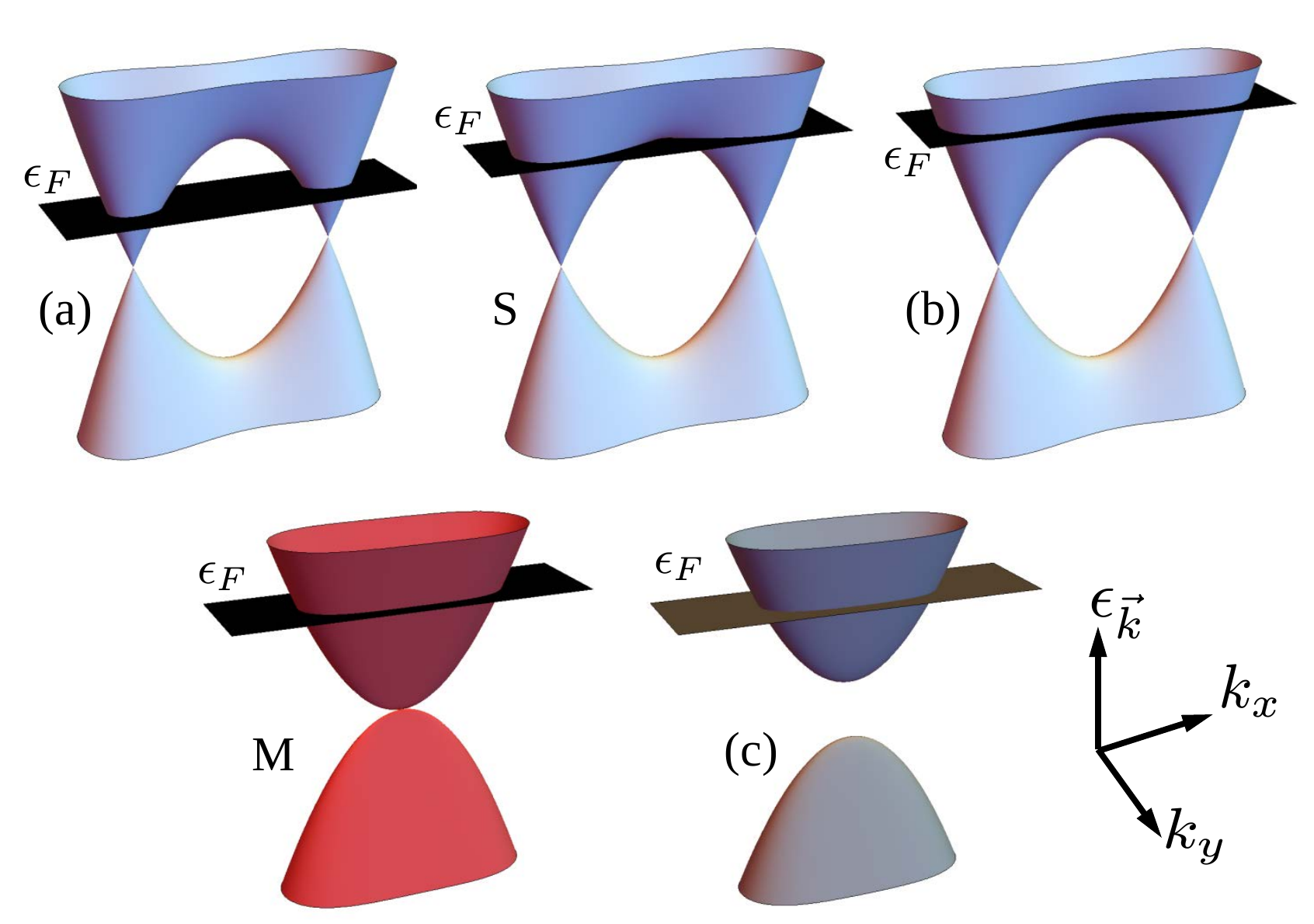}
\caption{This work addresses the transport properties for  an electronic spectrum undergoing a topological merging transition as
 depicted in this figure, and commented in more details in Fig. \ref{fig:EnergySpectrum}.
 }
\label{fig:EnergySpectrum-3D}
\end{figure}

To go beyond and in order to account for a  structure which consists in  two Dirac points separated by a saddle point, one needs  an appropriate  low energy $2 \times 2$ Hamiltonian. Moreover, {such a description is mandatory in} situations where, by varying band parameters, the Dirac points can be moved in reciprocal space.
Since these Dirac points are characterized by opposite topological charges, {they can even annihilate each other} \cite{Hasegawa:06,Guinea:08,Pereira:09,Montambaux:09a,Montambaux:09b}
. This merging is therefore a topological transition. It has been
shown that, at the transition, the electronic dispersion is quite
unusual since it is quadratic in one direction and linear in the other direction (the direction of merging). This "semi-Dirac" \cite{Banerjee:09} spectrum has new properties intermediate between a Schr\"odinger and a Dirac spectrum. The vicinity of the topological transition can be described by the following Hamiltonian {in two dimensions}\cite{Montambaux:09a,Montambaux:09b}:
\be H= \left(\Delta + {p_x^2 \over 2 m} \right) \sigma_x + c_y p_y \sigma_y \ . \ee
It has been coined   "Universal Hamiltonian" since the merging scenario of two Dirac points related by time reversal symmetry is uniquely described by this Hamiltonian\cite{Montambaux:09a,Montambaux:09b}.
 The parameter $\Delta$ drives the transition ($\Delta=0$) between a semi-metallic phase ($\Delta<0$) with two Dirac points and a gapped phase ($\Delta>0$), see Figs. \ref{fig:EnergySpectrum-3D},\ref{fig:EnergySpectrum}. The evolution of several thermodynamic quantities like the specific heat and the Landau level spectrum has been studied in details \cite{Montambaux:09a,Montambaux:09b}.

  %%%%%%%%%%%%%%%%%%%%%%%%%%%%%%%%%%%%%%%%%
%%%%%%%%%%%%%%%%%%%%%%%%%%%%%%%%%%%%%%%%%

 In this paper we address the {\it evolution of the conductivity tensor across the merging transition}  (Figs.  \ref{fig:EnergySpectrum-3D},\ref{fig:EnergySpectrum}).
  A first objective of this work is to characterize the transport properties as a possible signature of the evolution of the underlying band structure.
 On a more fundamental perspective, an additional interest of this problem
 stands from two important ingredients in the description of diffusive transport.
 First, at low energy the electronic wave functions have a spinorial structure  which leads to   effective anisotropic scattering matrix elements (similar to the case of a scalar problem with anisotropic scattering due to a disorder potential with finite range).
  This leads to a transport scattering time  $\tau^{\tr}$ different from the elastic scattering time $\tau_e$ , as in graphene for point-like impurities where $\tau^{\tr}= 2 \tau_e$.
Second, the anisotropy of the dispersion relation leads to an additional complexity: the   scattering times become themselves anisotropic and depend on the direction of the applied electric field. We show that within the Green's function formalism this anisotropy manifests itself into a rather unusual matrix
structure of the self-energy. A comparison between a Boltzmann approach and a perturbative Green's function formalism allows for a detailed understanding of this
physics.

The outline of the paper is the following. In the next section, we
recall the model, {\it  i.e.} the Universal Hamiltonian with coupling
to impurities described by a point-like white noise potential. We
define a directional density of states and derive the angular dependence of the elastic scattering time. In section \ref{sect:Diffusive}, we use the Boltzmann equation to calculate the conductivity tensor.
As a result of the two important ingredients mentioned above, the conductivity along a direction $\alpha$  is not simply proportional to the  angular averaged squared velocity $\langle v_\alpha^2(\theta) \rangle$ because~:
(i) the elastic  scattering time has also an angular dependence   due to the angular anisotropy of the spectrum, so that one should consider the average $\langle v_\alpha^2(\theta)  \tau_e(\theta)\rangle$; (ii) since the matrix elements of the interaction get an angular dependence, it will lead to   transport times  different of the elastic time. These transport times depend  on the direction $\alpha$  and, to obtain the conductivity, we will have   to consider the average $\langle v_\alpha^2(\theta)  \tau_\alpha^{\tr}(\theta)\rangle$. These results obtained from Boltzmann equation are confirmed by a diagrammatic calculation presented in section \ref{sec:diagrammatic}. We discuss our results in the last section.

\section{The Model}

\subsection{Hamiltonian and Fermi surface parametrization}
\label{sec:HamPure}

 %%%%%%%%%%%%%%%%%%%%%%%%%%%%%%%%%%%%%%%%%
\begin{figure} [!h]
\centering
\includegraphics[width=8cm]{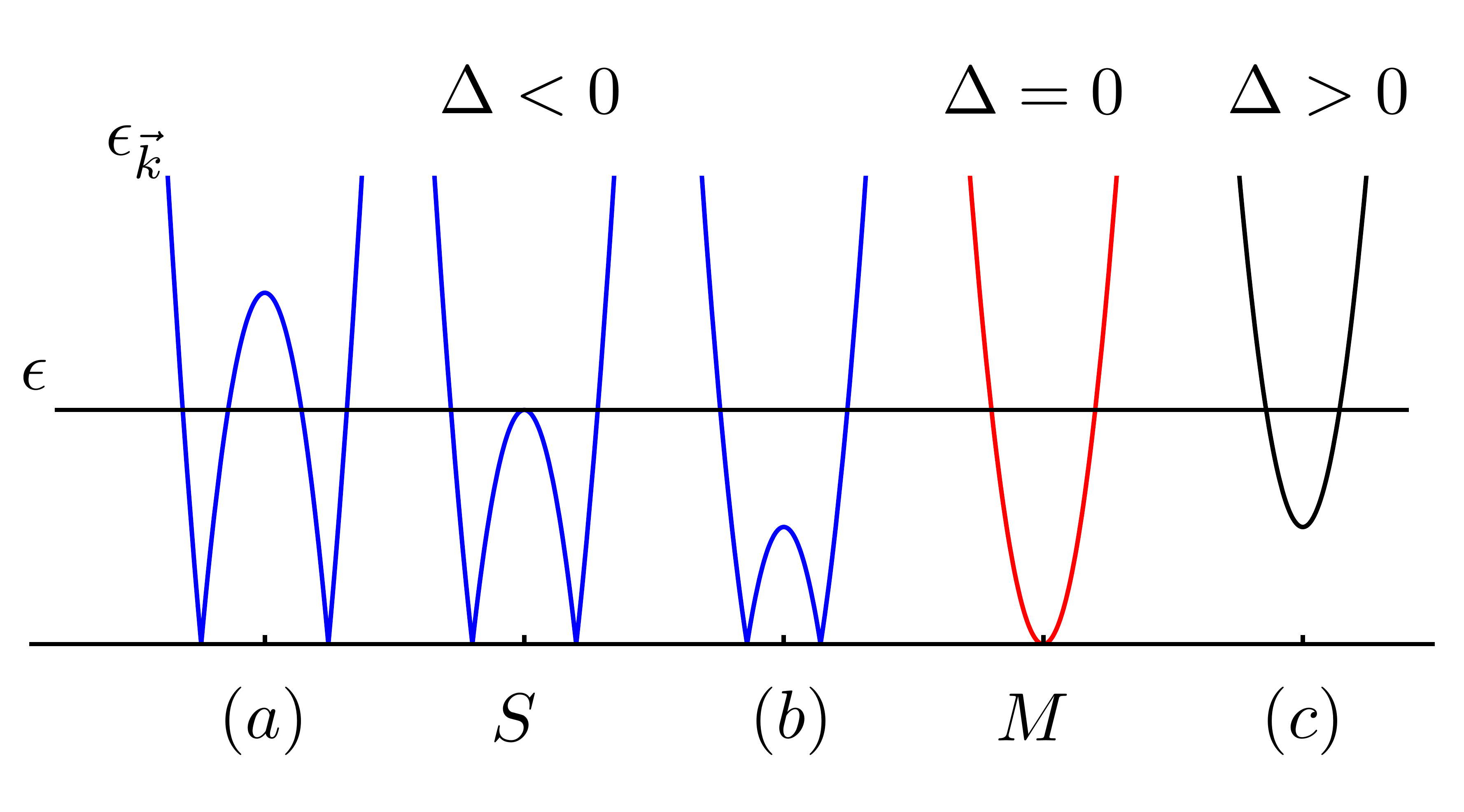}
\caption{Typical energy spectrum of the model (\ref{eq:PureModel})  for various  $\Delta $  but a fixed
energy $\epsilon >0$.
Dirac phase with
(a) $\Delta < - \epsilon$,
(S) $\Delta = - \epsilon $ (saddle-point) ,
(b) $ - \epsilon < \Delta <0$.
Critical semi-Dirac metal
(M) $\Delta = 0$.
Gapped phase
(c) $\Delta > 0$.
 }
\label{fig:EnergySpectrum}
\end{figure}
%%%%%%%%%%%%%%%%%%%%%%%%%%%%%%%%%%%%%%%%%

We consider the model described by the Hamiltonian
\begin{equation}
\label{eq:HModel}
 H=H^{0}+ V,
\end{equation}
where the disorder potential $V$ is defined and discussed in section \ref{sec:ScattTimes} and the Hamiltonian for the pure system is
defined as
\begin{equation}
\label{eq:PureModel}
H^{0} =   \left[ \Delta +  \frac{p_x^2}{2m}   \right] \sigma_x + c_y p_y \, \sigma_y \ .
\end{equation}
 In the present and the following sections (\ref{sec:HamPure} and \ref{sec:DensityStates})
we start by discussing  a few properties of the Hamiltonian $H^{0}$ without disorder.
 For $\Delta >0$ this Hamiltonian describes a gapped phase. When $\Delta < 0$, it  describes
 two Dirac cones with opposite chiralities, hereafter named a Dirac phase.
 Note that these Dirac cones are in general anisotropic with respectives velocities in the $x$ and $y$ directions  $c_x  = \sqrt{2 | \Delta | /m}$ and
$c_y$.
 The energy spectrum is given by
\begin{equation}
\epsilon^{2} =  \left( \frac{p_x^2}{2m} + \Delta \right)^{2} + \left( c_y p_y \right)^{2} \ .
\label{eq:EnergySpectrum}
\end{equation}
We will consider only the case of positive energies $\epsilon >0$, as the situation $\epsilon <0$ can be deduced from particle-hole symmetry.
Fig.~\ref{fig:EnergySpectrum} presents the different regimes discussed in this paper.

\begin{figure*}
\centering
\includegraphics[width=8cm]{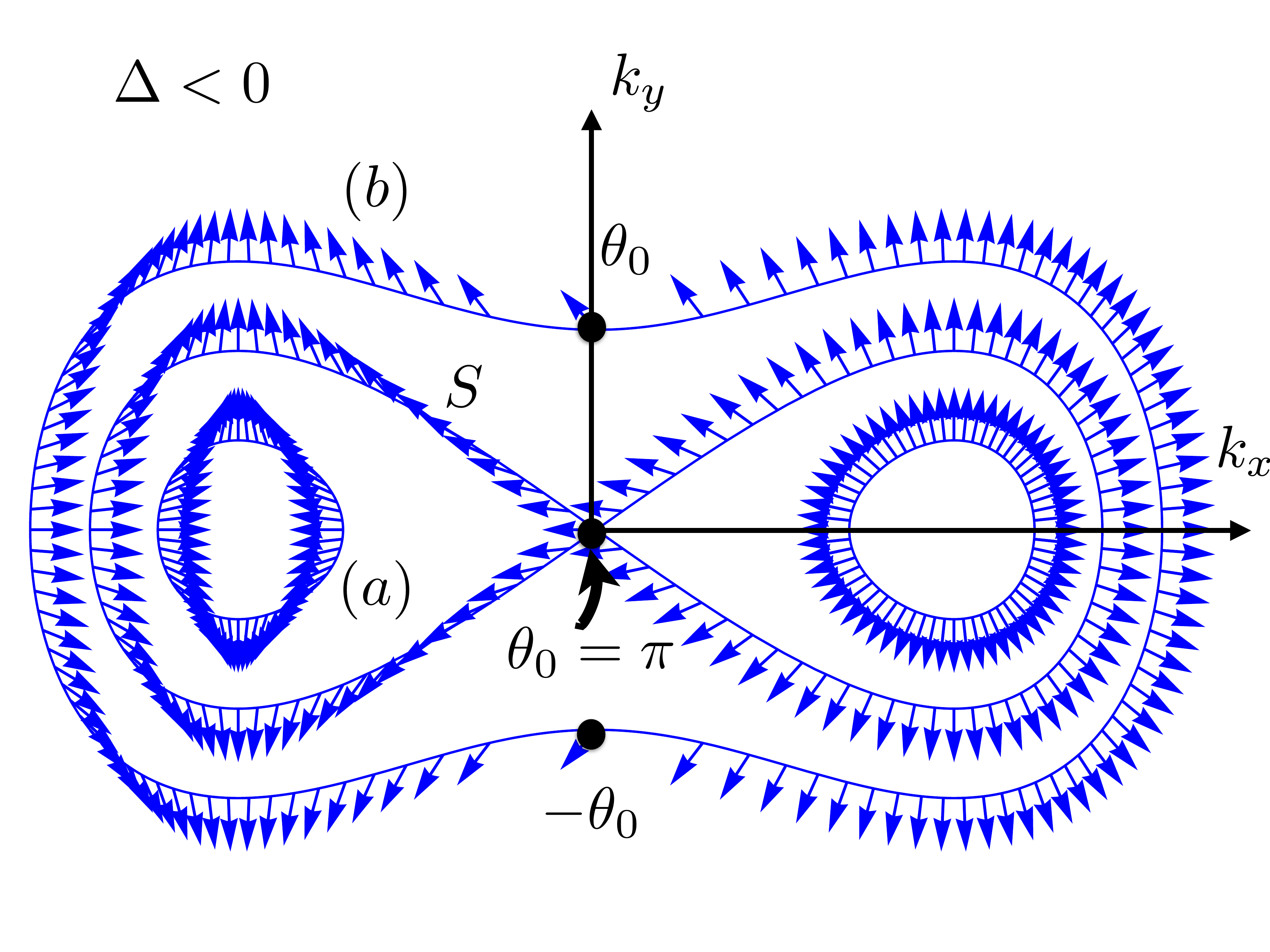}
\includegraphics[width=8cm]{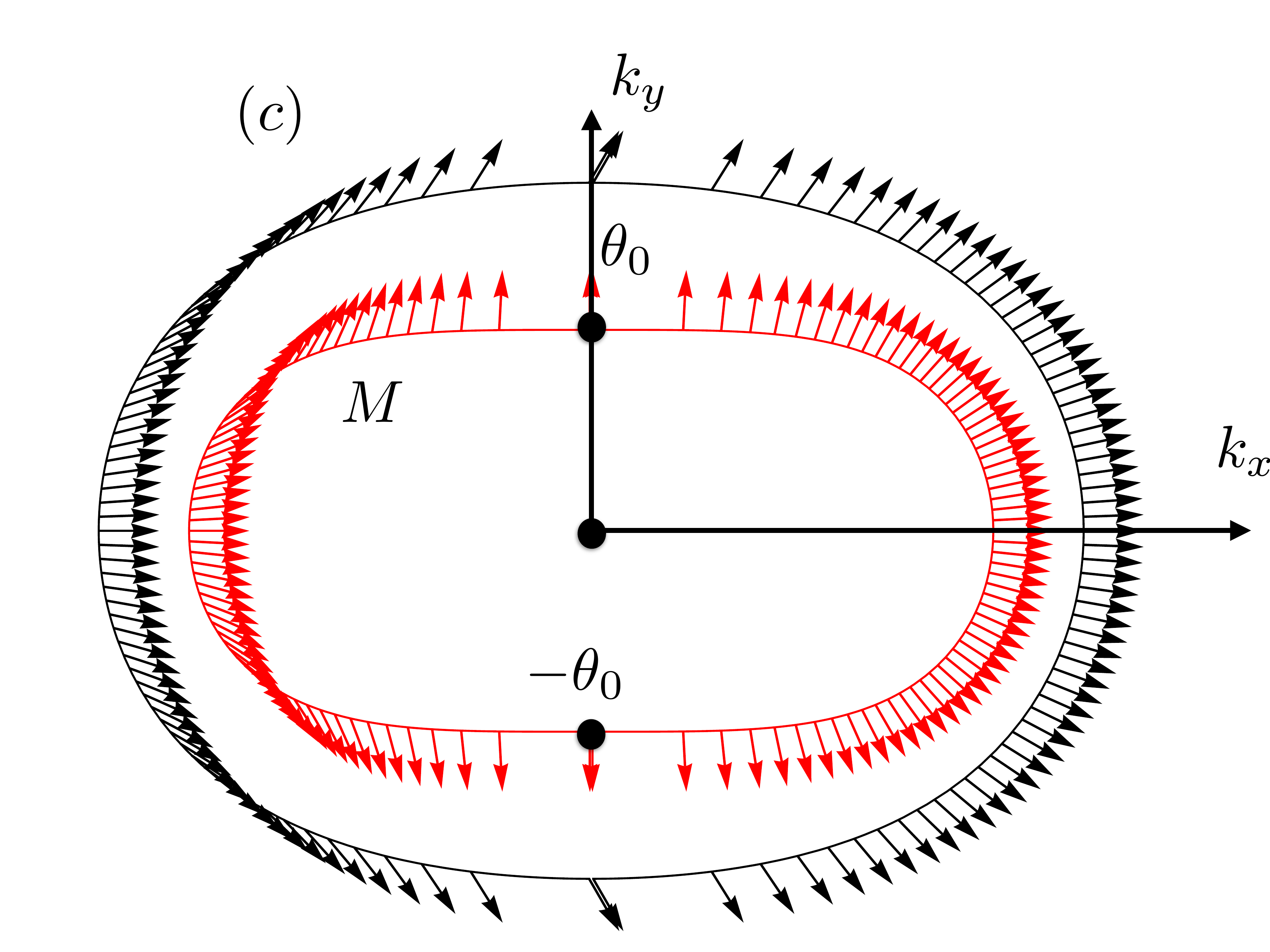}
\caption{Left~: Constant energy contours $\epsilon(k_{x},k_{y})$ for different energies and $\Delta <0 $
corresponding to the situations (a), (b) and the saddle-point S defined on Fig.~\ref{fig:EnergySpectrum}. The arrow field  describes the phase $\theta$ parametrizing  in a unambiguous way  each half $k_x>0$ and $k_x<0$ of
the energy contour according to Eq.~(\ref{eq:param}). It also describes the
 relative phase  between the two components of the eigenstate (\ref{eq:eigenstate}) of momentum $\vec{k}$ and energy $\epsilon$.   Right~: Same quantities  at the merging point  $\Delta=0$ (M) and for  $\epsilon(k_{x},k_{y}) > \Delta >0 \ (c)$.
 }
\label{fig:EnergyContour}
\end{figure*}

\begin{figure} [!h]
\centering
\includegraphics[width=8cm]{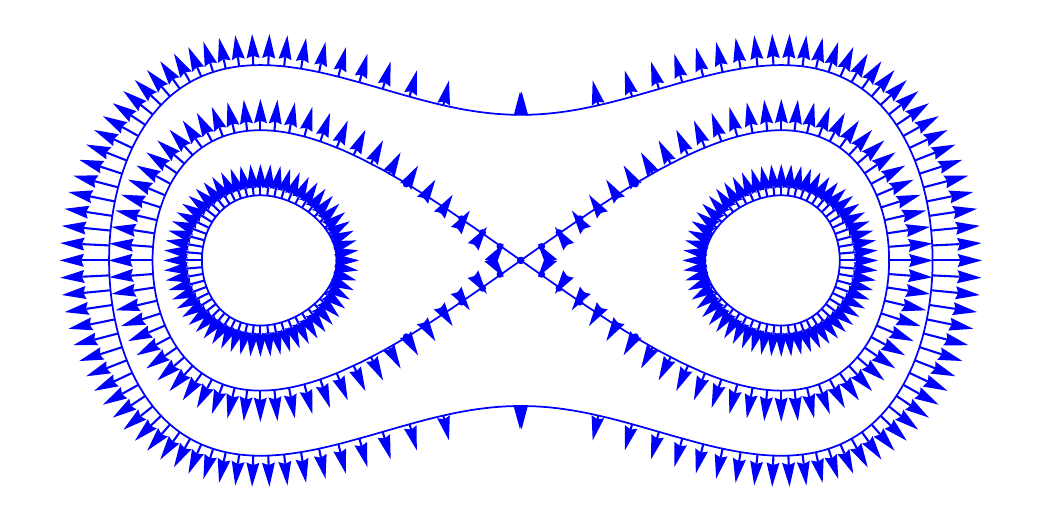}
\caption{Velocity $\vec v(\vec k)$ along constant energy contours (here $\Delta <0$).
}
\label{fig:vitesse}
\end{figure}

Parametrization of the constant energy contours of Eq.~(\ref{eq:EnergySpectrum}) is done by taking advantage of the $p_{x}$ parity.
For each half plane $p_{x} \lessgtr 0$ we use the parametrization
\begin{equation}
\frac{p_x^2}{2m} + \Delta = \epsilon \cos \theta
 \ ; \
c_y p_y = \epsilon \sin \theta
\  ; \
 {
\eta_{p}=\textrm{sign}(p_{x})=\pm
}
\label{eq:param}
\end{equation}
where $\theta \in [-\theta_{0},\theta_{0}]$  is a coordinate along the  constant energy contour.
{Its range depends on the topology of the constant energy contour,
and thus on the energy $\epsilon$, see Fig.~\ref{fig:EnergyContour}. Specifying the discussion to the Fermi surface associated with the Fermi energy $\epsilon_{F}$,
we can distinguish two cases :
\begin{itemize}
\item[(i)] Low energy metal with two disconnected  Fermi surfaces when $\Delta <0$ and $\epsilon_F <  - \Delta$. In this case $\theta_0 = \pi$.
This corresponds to the energy spectrum $(a)$ of Fig.~\ref{fig:EnergySpectrum}.
\item[(ii)] High energy metal with a single connected Fermi surface  for $\epsilon_F  >  | \Delta | $. In this case $\cos \theta_0 =  \Delta/\epsilon_F $.
For $\Delta <0$,
$\theta_0$ varies from $\pi$ for $\epsilon_F = -\Delta $ to $\pi /2$ for $\epsilon_F \gg  -\Delta$.
 For $\Delta >0$,  $\theta_0$ varies from $\pi/2$ for  $\epsilon_F \gg  \Delta$ to $0$
  for $\epsilon_{F} \to \Delta$ .
This corresponds to the energy spectra $(b),M,(c)$ of Fig.~\ref{fig:EnergySpectrum}.
\end{itemize}
}%
%%%%%%%%%%%%%%%%%%%%%%%%%%%%%%%%%%%%%%%%%
%%%%%%%%%%%%%%%%%%%%%%%%%%%%%%%%%%%%%%%%%
%
The eigenstates of positive energy corresponds to  wave functions
conveniently expressed with the parametrization of the constant energy contour
\begin{equation}
 \psi_{\vec{k}}(\vec{r})  =  \frac{1}{\sqrt{2}L} \left( \begin{array}{c} 1 \\ e^{i \theta_{\vec{k}}} \end{array} \right)  e^{i\vec{k}.\vec{r}} \ ,
 \label{eq:eigenstate}
\end{equation}
where $\theta_{\vec{k}}$ is defined by inversion of Eq.~(\ref{eq:param}), and $\vec{p} = \hbar \vec{k}$.
 From now on, we will set $L=1$.
The group velocity varies along the constant energy contour according to~:
\begin{subequations}
\begin{align}
v_{x} (\eta_p,\theta) & = \eta_p \sqrt{\frac{2\epsilon}{m}} \cos \theta \sqrt{\cos \theta - \delta } \ ,
\\
v_{y} (\eta_p,\theta) & = c_y \sin \theta \ ,
\end{align}
\label{eq:vitesse}
\end{subequations}
where throughout this paper we use the reduced parameter $\delta = \Delta /\epsilon$.
 The {evolution of the }velocity along constant energy contours is shown on Fig.~\ref{fig:vitesse}.

%%%%%%%%%%%%%%%%%%%%%%%%%%%%%%%%%%%%%%%%
%%%%%%%%%%%%%%%%%%%%%%%%%%%%%%%%%%%%%%%%%

\subsection{Density of States}
\label{sec:DensityStates}
We   define a  directional density of states along the constant energy contour parametrized
by $\theta$ from the equality
\begin{equation}
 \int \frac{dp_x dp_y}{(2 \pi \hbar)^2}=
  \int \frac{dk_x dk_y}{(2 \pi)^2} = \int \rho(\epsilon,\theta) ~d\epsilon d\theta \ ,
\end{equation}
with
{
 \begin{equation}
\rho(\epsilon,\theta)  =
\frac{\sqrt{2 m \epsilon}}{(2 \pi \hbar)^2 c_y} ~
\frac{1}{2 \sqrt{\cos \theta - \delta}} \ .
 \label{eq:dos12}
 \end{equation}
}%%%%
 The density of states in then obtained by the integral
\begin{equation}
\rho(\epsilon) =  2 \int_{-\theta_{0}}^{\theta_{0}} d\theta ~\rho(\epsilon,\theta) \,
\end{equation}
where the extra factor   $2$ accounts for   the sign of $p_x$.
The integral gives
\begin{equation}
\rho(\epsilon)  =    \frac{\sqrt{2 m \epsilon}}{(2 \pi \hbar)^2 c_y} ~   {I}_1(\delta) \ ,
\label{eq:defRho}
\end{equation}
with the function
\begin{equation}
{I}_1 (\delta)= \int_{-\theta_0}^{\theta_0}  {d \theta \over \sqrt{\cos \theta - \delta} } \ ,
\label{I1}
 \end{equation}
where $\theta_0 = \Arccos (\delta)$ when $|\delta|<1$ and $\theta_0 = \pi$ otherwise.
{
From Eqs.~(\ref{eq:dos12},\ref{eq:defRho}), we can rewrite $\rho(\ep,\theta)$ as
\begin{equation}
\label{eq:DirectionalDensityRewriting}
\rho(\ep,\theta) = {\rho(\ep) \over 2 \,   I_1(\delta) \sqrt{\cos \theta - \delta}}  \ .
 \end{equation}
}%%%%

{
\subsection{Disorder Potential and Elastic Scattering Time}
\label{sec:ScattTimes}
}

The disorder part $V$ of the Hamiltonian accounts for the inhomogeneities in the system.
This random potential $V(\r)$  is assumed to describe a gaussian
  point-like uncorrelated disorder, characterized by   two cumulants
\begin{equation}
\overline{V(\r)}  = 0,
\quad
\overline{V(\vec{r})V(\vec{r'}) } = \gamDes \, \delta(\vec{r}-\vec{r'}) \ .
\end{equation}
{where the overline denotes a statistical average over realizations of the random potential.}
The presence of  this random potential induces a finite lifetime for the eigenstates of momentum $\k$
of the pure model (\ref{eq:PureModel}), {called elastic scattering time, and
}%%%%
obtained from the Fermi  golden rule~:
\begin{align}
\frac{\hbar}{\tau_{e}(\vec{k})} &=
2\pi \int  \frac{d^2 \vec{k}'}{(2\pi)^2}
\delta (\epsilon_{\vec{k}} - \epsilon_{\vec{k}'} )
\overline{|\mathcal{A}(\vec{k},\vec{k}')|^2} \  ,
\end{align}
where the scattering amplitude is defined by
\begin{equation}
\mathcal{A}(\vec{k},\vec{k}') = \langle \psi_{\vec{k}} |V| \psi_{\vec{k}'}\rangle  \ .
\end{equation}
For uncorrelated point-like disorder, the angular dependence of this scattering amplitude originates from the eigenstates overlap and one has
\begin{equation}
\label{eq:scattamplitude}
\overline{|\mathcal{A}(\vec{k},\vec{k}') |^2 } =
 \frac{\gamDes}{2}    \left( 1 + \cos (\theta_{\vec{k}}-\theta_{\vec{k}'}) \right) \ .
\end{equation}
%
%
%\begin{equation}
%\frac{1}{\tau_{e}(\vec{k})} = \frac{2 \pi}{\hbar} \oint_{\mathcal{C}_{\epsilon}} d\theta \rho(\epsilon, \theta)
%|\langle \psi_k |V| \psi_k'\rangle |^2  ~ .
%\end{equation}
%
Defining $\tau_{e}(\epsilon,\theta)=\tau_e(\vec{k})$ where $\epsilon, \theta$ and $\vec{k}$ are related through
Eq.~(\ref{eq:param}), we can express the elastic scattering time as an  integral
\begin{equation}
 \label{eq:tauepstheta}
\frac{\hbar}{\tau_{e}(\epsilon,\theta)}
=
2 \pi \gamDes  \int_{-\theta_0}^{\theta_0} d\theta'~\rho(\epsilon,\theta')
 \left[ 1 + \cos (\theta-\theta') \right] .
\end{equation}
{
Introducing the bare scattering time
\begin{equation}
 \tau_{e}^{0}  (\epsilon) = \frac{ \hbar }{ \pi  \gamDes \rho({\epsilon })}  \ ,
\label{eq:ElasticTimeMean}
 \end{equation}
we can rewrite (\ref{eq:tauepstheta}) in the form
\begin{equation}
\tau_{e}(\epsilon,\theta)
 =\frac{ \tau_{e}^{0}  (\epsilon) }{ 1  + r(\delta ) \cos \theta } \ ,
\label{eq:ElasticTimeParam}
 \end{equation}
 where the density of states $\rho(\epsilon)$ is given by (\ref{eq:defRho}).
 The denominator of this expression exactly accounts for the anisotropy of the scattering time. As a convenient parametrization of this property, we }
 have introduced the  anisotropy function $r(\delta)$ which will be used throughout this
paper :
\begin{equation}
r(\delta) =   {J}_1 \left(\delta \right)  /  {I}_1\left(\delta \right),
\label{eq:defR}
\end{equation}
with the function $I_1(\delta)$ defined in (\ref{I1})
and
\begin{equation}
 {J}_1 (\delta)=  \int_{-\theta_0}^{\theta_0}  {d \theta \cos \theta \over \sqrt{\cos \theta - \delta} } \ ,
\label{J1}
 \end{equation}
where $\theta_0 = \Arccos (\delta)$ when $|\delta|<1$ and $\theta_0 = \pi$ otherwise.
%
%%%%%%%%%%%%%%%%%%%%%%%%%%%%%%%%%%%%%%%%%
\begin{figure} [!h]
\centering
\includegraphics[width=8cm]{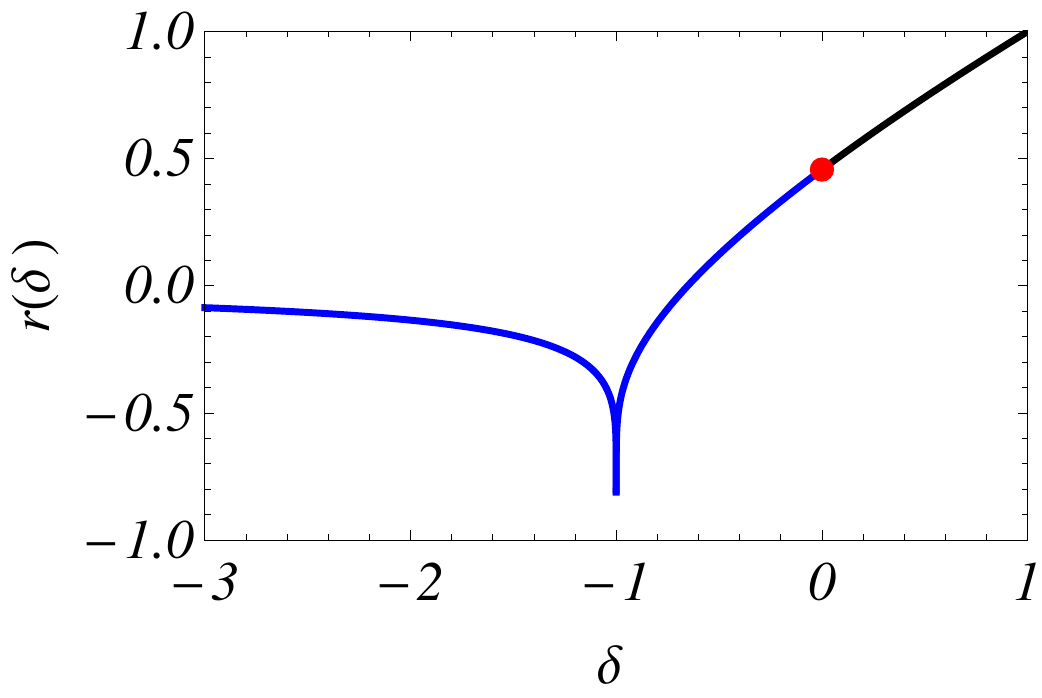}
\caption{Function $r(\delta)$ parametrizing the angular dependence of the elastic scattering time $\tau_e$ plotted as a
function of $\delta = \Delta / \epsilon$.
It has the limits
$r(\delta \rightarrow - \infty) \simeq 1/(4\delta)$, $r(-1)=-1$, $r(0)= 2 \Gamma(3/4)^4 /\pi^2 \simeq 0.456947$, $r(1)=1$.
In this figure, as in following figures, we systematically reserve the colors~: blue for the Dirac phase ($\delta < 0$), black for the gapped phase ($\Delta > 0$) and red for the semi-Dirac point.
 }
\label{fig:r}
\end{figure}
%%%%%%%%%%%%%%%%%%%%%%%%%%%%%%%%%%%%%%%%%
  The function $r(\delta)$
is plotted in Fig.~\ref{fig:r}.  Deep in the Dirac phase ($\Delta \ll 0$), at low energy ($\ep \ll |\Delta|$), one has
\begin{equation}
r(\delta)   \rightarrow   -{ 1 \over 4 |\delta|}=-{ \epsilon \over 4 |\Delta|} \ll 1 ,
 \end{equation}
so that the anisotropy  can be neglected in Eq.~(\ref{eq:ElasticTimeParam}) and we recover a scattering time independent of the direction of
propagation as standard for Dirac fermions.

%%%%%%%%%%%%%%%%%%%%%%%%%%%%%%%%%%%%%%%%%
%\begin{figure*} [!h]
%\centering
%\includegraphics[width=5cm]{Figures/Energy.pdf}
%\includegraphics[width=5cm]{Figures/Contour-D0.pdf}
%\\
%\includegraphics[width=5cm]{Figures/Contour-D05.pdf}
%\includegraphics[width=5cm]{Figures/Contour-D-05.pdf}
%\caption{Energy as a function of $p_{x}$ for $p_{y}=0$ and $\Delta=0, \Delta =\pm 0.5$. Shape of constant energy curves for the same parameters.  }
%\label{fig:atilde}
%\end{figure*}
%%%%%%%%%%%%%%%%%%%%%%%%%%%%%%%%%%%%%%%%%

\section{Diffusive regime from the Boltzmann Equation}
\label{sect:Diffusive}

We now consider the transport properties of the model (\ref{eq:HModel}) at a fixed   energy $\epsilon$  large enough so that the condition $k l_e \gg 1$ is fulfilled, $l_e$ being a typical elastic  mean free path. Therefore we will not consider the close vicinity of a Dirac point and the associated physics of minimal conductivity
\cite{Katsnelson:2006b,Twordzylo:2006}.
For a system of typical size much larger than this mean free path $l_e$, this corresponds to the regime  of classical diffusion.
We describe this regime first with  a standard Boltzmann equation, before turning to a complementary but equivalent diagrammatic approach based on
Kubo formula for the conductivity. The use of these two approaches will reveal the physics hidden between the technical specificities of the diffusive
transport for the model we consider.

\subsection{Boltzmann equation}

We start from the Boltzmann equation \cite{Abrikosov:88,Ziman:79} expressing the evolution of the distribution function $f(\vec{k},\vec{r})$ :
\begin{equation}
\frac{d f}{dt} + \frac{d \vec{r}}{dt} ~ \nabla_{\vec{r}}  f+ \frac{d \vec{k}}{dt} ~ \nabla_{\vec{k}} f =
I[f] \  ,
\label{eq:Boltzmann0}
\end{equation}
where $I[f]$ is the collision integral defined below.
The position $\vec{r}$ and momentum $\vec{k}$ parametrizing the distribution function $f(\vec{k},\vec{r})$ are classical variables, whose time evolutions
entering Eq.(\ref{eq:Boltzmann0}) are described by the semi-classical equations \cite{Xiao:2010,Son:2013}
\begin{align}
\frac{d \vec{r}}{dt} &= \vec{v}(\vec{k})  +
\frac{d \vec{k}}{dt} \times \vec{F}_{\vec{k}} \\
\hbar \frac{d \vec{k}}{dt} &= -e \vec{E} - e \frac{d \vec{r}}{dt} \times \vec{B}
\label{eq:semiclassic2}
\end{align}
with the group velocity
$ \vec{v}(\vec{k})  = \hbar^{-1}\partial \epsilon(\vec{k}) / \partial \vec{k}$,
$\vec{B}$ is a local magnetic field, and
$\vec{F}_{\vec{k}} =  i  \nabla_{\vec{k}}\times  \langle \psi_{\vec{k}} |\nabla_{\vec{k}} \psi_{\vec{k}} \rangle$
 is the Berry curvature.
In the present case, we consider the response of the distribution function
 $f(\vec{k},\vec{r})$  due to a uniform weak electric field $\vec{E}$ : we can neglect the gradient $ \nabla_{\vec{r}}  f$
  in Eq.~(\ref{eq:Boltzmann0}) and drop the spatial dependence of $f$.
  Due to the absence of magnetic field,  we deduce from  Eqs.~(\ref{eq:Boltzmann0},\ref{eq:semiclassic2}) that
  a stationary out-of-equilibrium distribution $f(\vec{k})$  satisfies the simpler equation
\begin{equation}
- { e \over \hbar } \vec{E} .\nabla_{\vec{k}} f =   I[f] \ .
 \label{eq:Boltzmann}
\end{equation}
where $f$ is now as function of $\vec{k}$ and the collision integral is expressed as
\begin{multline}
I[f] =
2\pi \int \frac{d^2 \vec{k}'}{(2\pi)^2} \\
 \delta (\epsilon_{\vec{k}} - \epsilon_{\vec{k}'} )
  \overline{ |\mathcal{A}(\vec{k},\vec{k}') |^2 }
\left( f(\vec{k}') - f(\vec{k}) \right)   .
\label{eq:Idef}
\end{multline}

By assuming the perturbation to be weak, we can expand the stationary out-of-equilibrium distribution $f(\vec{k})$
around the equilibrium Fermi distribution $f^0(\vec{k}) = n_{F}(\epsilon_{\vec{k}})$ following
the ansatz\cite{Abrikosov:88,Ziman:79}
\begin{equation}
f(\vec{k}) = f^0(\vec{k})  + e \frac{\partial n_F }{\partial \epsilon } ~ \vec{\Lambda}(\vec{k}).\vec{E} \ ,
\label{eq:BoltzAnsatz}
\end{equation}
where the vector $\vec{\Lambda}$ has the dimension of a  length, and its components correspond  to transport lengths in the different spatial directions.
They are related to transport times through the definition
$\Lambda_{\alpha}(\vec{k}) =  v_{\alpha}(\vec{k})  \tau^{\textrm{tr}}_{\alpha} (\vec{k})  $.
Eq.~(\ref{eq:BoltzAnsatz}) can be rewritten as a shift of energies by the field~:
$f(\vec{k}) = n_{F}\left(\epsilon_{\vec{k}} + e \vec{\Lambda}(\vec{k}).\vec{E} \right)$. In the case of an isotropic Fermi surface, we do not expect this
shift to depend on the direction of application of the field $\vec{E}$ : in that case a unique transport time $\tau^{\textrm{tr}}$ is necessary to
describe the stationary distribution\cite{Ziman:79}.
Here,  for an anisotropic Fermi surface such as (\ref{eq:EnergySpectrum}), we generically expect the response of the distribution
function to
{depend on the direction of the electric field $\vec{E}$ \cite{Sondheimer:1962,Sorbello:1974,Sorbello:1975}.
For an electric field applied in the $x$ or $y$ direction, this leads}
to the definition of different anisotropic transport times $\tau^{\textrm{tr}}_x,\tau^{\textrm{tr}}_y$.
 From    Eqs. (\ref{eq:Boltzmann},\ref{eq:Idef},\ref{eq:BoltzAnsatz}), one obtains
\begin{widetext}
%\begin{multline}
\begin{equation}
\vec{v}(\vec{k})  =
\left. \frac{1}{\hbar}\frac{\partial \epsilon}{\partial \vec{k}}\right|_{\epsilon=\epsilon_{\vec{k}}}
%\\
=
2\pi \int \frac{d^2 \vec{k}'}{(2\pi)^2}
 \delta (\epsilon_{\vec{k}} - \epsilon_{\vec{k}'} )
  \overline{ |\mathcal{A}(\vec{k},\vec{k}') |^2 }
\left(  \vec{\Lambda}(\vec{k}) -  \vec{\Lambda}(\vec{k}') \right) \ ,
\end{equation}
%\end{multline}
%

By using the parametrization (\ref{eq:param}) on the contour of constant energy $\ep$, each component $\alpha$ of the velocity
obeys the equation (to lighten notation, we omit the energy $\ep$ in the argument of the quantities in the next expressions)~:
% and assuming that
% $\Lambda_{\alpha}(\vec{k})$ possesses the same symmetries as $v_{\alpha}(\vec{k})$
%
%\begin{multline}
\begin{equation}
v_\alpha(\eta_p,\theta)  = {\Lambda_\alpha(\eta_p,\theta) \over \tau_e(\theta)}
%\\
- \frac{  \pi \gamDes}{\hbar} \sum_{\eta'_p=\pm} \int_{-\theta_0}^{\theta_0} d\theta' ~ \rho( \theta')
\left[1+\cos(\theta - \theta') \right]
   \Lambda_\alpha(\eta'_p,\theta')   .
\end{equation}
%\end{multline}
%
The transport times $\tau^{\textrm{tr}}_{\alpha} (\epsilon, \theta)$ are defined   as
\begin{equation}
\Lambda_{\alpha}(\ep, \eta_p,\theta) =  v_{\alpha}(\ep,  \eta_p,\theta) \tau^{\textrm{tr}}_{\alpha} (\ep, \theta) \ .
\end{equation}

We now assume the following ansatz, namely that the transport times and the elastic scattering time have the same angular dependence:
\begin{equation}
  \tau^{\textrm{tr}}_{ \alpha} (\ep, \theta) = \lambda_\alpha(\ep) \,  \tau_e (\ep, \theta)  \ ,
\end{equation}
so that the parameters $\lambda_\alpha(\ep)$ are obtained  from  the self-consistent equation (at fixed energy $\ep$)
%\begin{multline}
\begin{equation}
v_\alpha (\eta_p,\theta)= \lambda_\alpha  v_\alpha (\eta_p,\theta)
%  \\
- \frac{ \pi \gamDes}{\hbar}\lambda_\alpha \sum_{\eta'_p=\pm} \int_{-\theta_0}^{\theta_0} d\theta' \rho( \theta')
\left[1+\cos(\theta - \theta') \right]
 v_\alpha(\eta'_p,\theta')  \tau_e(\theta')
 \label{eq:Lambda}
\end{equation}
%\end{multline}
where $v_\alpha( \eta_p,\theta)$ is defined in Eq.~(\ref{eq:vitesse}).
Then from Eq.~{  (\ref{eq:DirectionalDensityRewriting})} and (\ref{eq:ElasticTimeParam}), we finally get
%
%\begin{multline}
\begin{equation}
v_\alpha (\eta_p,\theta)= \lambda_\alpha  v_\alpha (\eta_p,\theta)
%  \\
-  {\lambda_\alpha \over 2 I_1(\delta) }\sum_{\eta'_p=\pm} \int_{-\theta_0}^{\theta_0} d\theta'
{1+\cos(\theta - \theta')  \over  1 +r(\delta) \cos \theta'}
 {v_\alpha(\eta'_p, \theta') \over \sqrt{\cos \theta'  - \delta}}  \ .
 \label{eq:lambda}
\end{equation}
%\end{multline}
\end{widetext}

We now consider the two directions $\alpha = x,y$ separately.

{\it Along the $x$ direction}, since the velocity is an odd  function of $k_x$,
the sum over $\eta_p$ in Eq.~(\ref{eq:lambda}) vanishes:  we obtain $\lambda_x(\ep)=1$, {\it i.e.} the transport time is
 equal to the scattering time\footnote{This is not true in graphene where $\lambda_x=2$. Note that this
peculiar result ($\lambda_x=1$) is due  the fact that   the matrix
elements of the disorder potential are supposed here to have no
momentum dependence. Assuming an opposite limit where the disorder
would not couple valleys, then in the Dirac limit $0 < \ep \ll -
\Delta$, one would recover  $\lambda_x=1$ and  $\tau_x^{\textrm{tr}}=
2 \tau_e$.}
{$  \tau^{\textrm{tr}}_{x} (\ep, \theta) =  \tau_e (\ep, \theta) $}.

\begin{figure} [!h]
\centering
\includegraphics[width=8cm]{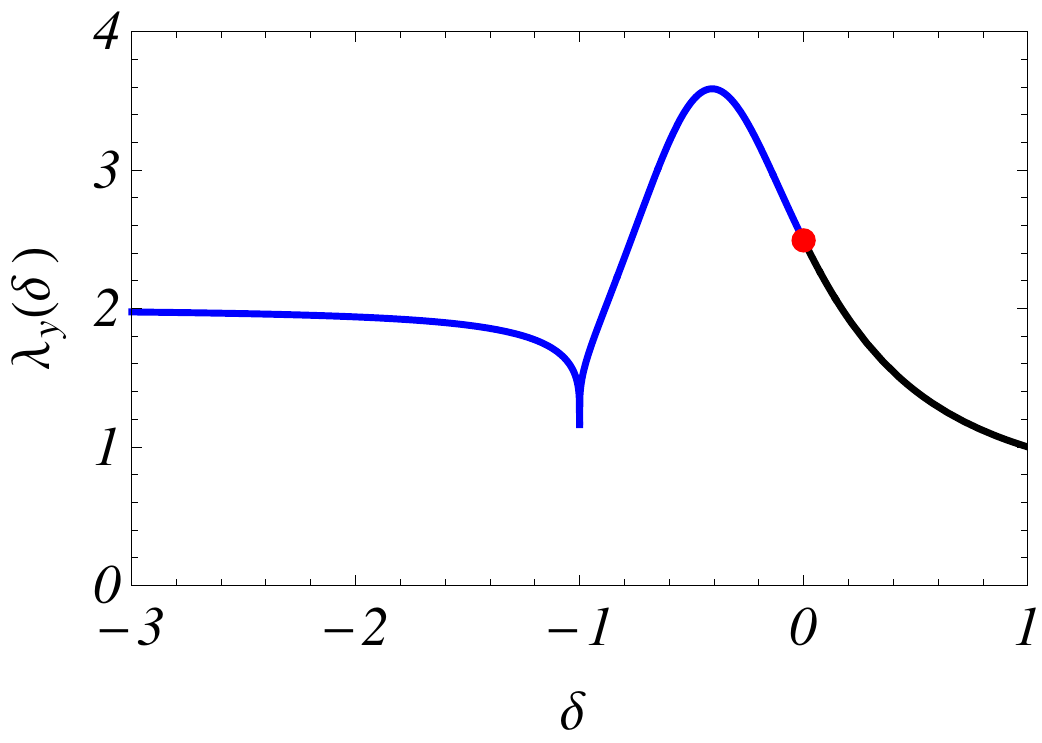}
\caption{
Dependence on $\delta = \Delta /\epsilon$ of the
renormalization factor  of the transport time  $\tau^{\textrm{tr}}_{ y}$ with respect to the elastic scattering time :
$\lambda_y(\delta)= \tau^{\textrm{tr}}_{ y}({\k}) / \tau_{e}({\k})$.
}
\label{fig:lambday}
\end{figure}

{\it Along the $y$ direction}, where $v_y(\theta)= c_y \sin \theta$ independent of $\eta_p$,
Eq.~(\ref{eq:lambda}) possesses a self-consistent solution, and we obtain
\begin{equation}
 \lambda_y(\delta) = \frac{1}{1 -  \mathcal{I}_2( \delta )/   {I}_1( \delta )}
 \label{eq:lambday}
\end{equation}
where
\begin{equation}
 \mathcal{I}_2(\delta) =
\int_{-\theta_{0}}^{\theta_{0}} d \theta ~ \frac{\sin^2 \theta}{\sqrt{\cos \theta - \delta}~ (1 +r(\delta) \cos \theta)} ~ .
\label{eq:calI2}
\end{equation}
The function $I_1(\delta)$ is defined in (\ref{I1}). Note that the expression (\ref{eq:lambday}) of the renormalization factor of
the transport time $\lambda_y(\delta)$
reflects the iterative structure
of the vertex correction to the bare conductivity that will be obtained within
a diagrammatic treatment in section \ref{sec:DiagConductivity} (see Eqs.~\ref{eq:RenormJy},\ref{eq:sigma_yy_diag}). The dependence $\lambda_y(\delta)$ is plotted in Fig.~\ref{fig:lambday}.

%%%%%%%%%%%%%%%%%%%%%%%%%%%%%%%%%%%%%%%%
%%%%%%%%%%%%%%%%%%%%%%%%%%%%%%%%%%%%%%%%%

Having obtained the transport times along the $x$ and $y$ directions, we now turn to the calculation of the conductivities.

\subsection{Conductivity}
\label{sect:conductivity}

We can express the current density $\vec{j}$ occurring in response to the application of the electric field $\vec{E}$ as
\begin{equation}
\vec{j}  =
\int \frac{d^2 \vec{k}}{(2\pi)^2}
\left[f(\vec{k}) - n_F (\epsilon_{\vec{k}})\right] (-e \vec{v}(\vec{k}))  \ .
\end{equation}
By using $\partial n_F / \partial \epsilon \simeq - \delta(\epsilon-\epsilon_{F})$
and the ansatz (\ref{eq:BoltzAnsatz}) for the distribution function $f(\vec{ k})$  we obtain
\begin{multline}
\vec{j}  =
e^2 \\ \sum_{\eta_p=\pm}
 \int_{-\theta_0}^{\theta_0} d\theta ~  \rho(\epsilon_{F},\theta)
\vec{v}(\epsilon_{F},\eta_p, \theta) \left[  \vec{\Lambda}(\epsilon_{F},\eta_p,\theta).\vec{E} \right] \ .
\end{multline}
The symmetries of this equation imply that off-diagonal terms of the conductivity tensor vanish
($\sigma_{\alpha, \beta\neq \alpha} = 0$) while the diagonal terms  can be written as
\begin{equation}
\sigma_{\alpha \alpha} =
2 e^2
 \int_{-\theta_0}^{\theta_0} d\theta ~ \rho(\epsilon_{F},\theta)
v_{\alpha}(\epsilon_{F},\theta)   \Lambda_{\alpha}(\epsilon_{F},\theta)  \ .
\end{equation}
where the factor $2$ originates from the two possible signs of $\eta_p=\pm $.
We end up with the Einstein relation
{
\begin{equation}
\sigma_{\alpha\alpha} = e^{2} \rho(\epsilon_{F}) D_{\alpha} ,
\end{equation}
with the diffusion coefficients
\begin{subequations}
\begin{align}
D_{\alpha } &=
2 \lambda_\alpha(\ep_F)
 \int_{-\theta_0}^{\theta_0} d\theta ~ { \rho(\epsilon_{F},\theta) \over \rho(\epsilon_{F})}
 v^2_{\alpha}(\theta)  \tau_e(\theta) \\
&=
 \left<  v^2_{\alpha}(\epsilon_{F},\theta) ~ \tau^{\textrm{tr}}_{\alpha}(\epsilon_{F},\theta)  \right>_{\theta}
 \\
 &=
 \lambda_\alpha(\ep_F)
 \left<  v^2_{\alpha}(\epsilon_{F},\theta)  \tau_e(\epsilon_{F},\theta)  \right>_{\theta}
 .
\end{align}
\end{subequations}
where we have defined the average along the constant energy contour
$\left< \cdots  \right>_{\theta} = 2
\int_{-\theta_0}^{\theta_0} d\theta ~ \cdots   \rho(\epsilon_{F},\theta) / \rho(\epsilon_{F})
$.
 This corresponds to the result announced in the introduction~: the diffusion coefficients $D_\alpha$  are obtained by an average over the Fermi surface
 of $v^2_{\alpha}~ \tau^{\textrm{tr}}_{\alpha}$ instead of $v^2_{\alpha}~  \tau_e$. With our solution of the Boltzmann equation, this
 difference is accounted for by a renormalization factor $\lambda_\alpha(\ep_F)$ of the diffusion coefficients, which does not depend
 on the direction along the Fermi surface but \emph{on the direction $\alpha$ of application of the electric field}.
}%%%%
 We now specify explicitly the conductivities  along the two directions $x$ and $y$.

{\it Along the $x$ direction}, there is no  renormalization of the transport time ($\lambda_x=1$, $\tau_x^{\textrm{tr}}=\tau_e$) and the     conductivity $\sigma_{xx}$ reads
\begin{align}
\sigma_{xx} &  =
2 e^2 \int_{-\theta_{0}}^{\theta_{0}} d\theta ~
 \rho(\epsilon_{F},\theta) v_x^2(\theta) \tau_e(\theta) \nonumber \\
&  =
 \frac{ e^2  \hbar}{ \pi \gamDes} ~
 \frac{2 \epsilon  }{m  } ~
 \frac{ \mathcal{I}_3(\Delta/\epsilon)}{  {I}_1(\Delta/\epsilon)} \ ,
\label{eq:sigma_xx_boltzmann}
\end{align}
where we define
\begin{equation}
 \mathcal{I}_3(\delta)=
 \int_{-\theta_{0}}^{\theta_{0}} d \theta ~ \frac{\cos^2 \theta \sqrt{\cos \theta -\delta }}{1 + r(\delta) \cos \theta} \ ,
\end{equation}
and  the function $I_1(\delta)$ is given in (\ref{I1}).

 For the conductivity {\it along the $y$ direction}, the renormalization of the transport time is given by  (\ref{eq:lambday}) and   we obtain
\begin{align}
\sigma_{yy} & =
2 e^2  \lambda_y(\Delta /\epsilon)
 \int_{-\theta_{0}}^{\theta_{0}} d\theta ~
 \rho(\epsilon_{F},\theta) v_y^2(\theta) \tau_e(\theta) \nonumber \\
&  =
 \frac{ e^2  \hbar}{ \pi \gamDes} ~ c_y^2 ~
 \frac{ \mathcal{I}_2(\Delta/\epsilon)}{I_1(\Delta/\epsilon) -  \mathcal{I}_2(\Delta/\epsilon) } \ ,
 \label{eq:sigma_yy_boltzmann}
\end{align}
where the functions $I_1(\delta)$ and $\mathcal{I}_2(\delta)$ are respectively given by Eqs. (\ref{I1}) and (\ref{eq:calI2}).
Eqs.~(\ref{eq:sigma_xx_boltzmann}, \ref{eq:sigma_yy_boltzmann}) constitute the main results of this work.
We discuss them in section \ref{sect:discussion}.
In the next section, we use a diagrammatic approach
{
which proposes a complementary description of the anisotropy of transport and allows to
 confirm the ansatz made to solve the Boltzman equation and recover exactly the  results of Eqs.~(\ref{eq:sigma_xx_boltzmann}, \ref{eq:sigma_yy_boltzmann})
 .
}%%%%

\section{Diagrammatic Approach}
\label{sec:diagrammatic}

{

An alternative approach to describe the diffusive transport of
electron consists in a perturbative expansion in disorder of the
conductivity tensor using a diagrammatic technique\cite{Akkermans}.
Beyond confirming the ansatz made to solve the Boltzmann equation
described above, this method allows for an instructive alternative
treatment of the different transport anisotropies.
In the diagrammatic approach, the  transport coefficients of the model
are obtained from the Kubo formula. A perturbative expansion is then
used to express the transport coefficients using the average single particle
Green's function.
In this formalism, the
anisotropy of scattering and transport times are cast into a unusual
matrix form for the self-energy operator $\Sigma$.
Beyond the present model, such a technique allows to
describe anisotropy of diffusion of Dirac fermion models due {\it
  e.g.}  to the warping of the Fermi surface in topological insulators
\cite{Adroguer:12} or anisotropic impurity scattering, the study of
which goes beyond the scope of the present paper.
 Nevertheless our work provides a physical understanding of the technicalities naturally occurring in these other problems.
In the next
subsections, we   first discuss the self-energy and the single
particle Green's function. We   then turn to the calculation of the
conductivity.

\subsection{Green's functions and self-energy}

The retarded and advanced Green's functions are defined by~:
\begin{multline}
G^{R/A}(\vec{k},\vec{k}', \epsilon_{F}) = \\
\left[
\left( (\epsilon_{F} \mp i 0)   \mathbf{I}  - H^0(\vec{k}) \right)\delta(\vec{k}-\vec{k}') - V(\vec{k},\vec{k}') \mathbf{I}
\right]^{-1}
\end{multline}
In the case of the model without disorder defined by Eq.~(\ref{eq:PureModel}), the  Green's function  is expressed as a $2 \times 2$ matrix~:
\begin{align}
G^0(\vec{k},\epsilon) &= \left( \epsilon \,  \mathbf{I}  - H^0(\vec{k}) \right)^{-1}
\nonumber \\
&=
\frac{\epsilon \, \mathbf{I} + \left( \frac{\hbar^2 k_x^2}{2m}+\Delta \right) \sigma_x  + c_y~ \hbar k_y \sigma_y }
{\epsilon^2  -\left( \frac{\hbar^2 k_x^2}{2m}+\Delta \right)^2- c_y^2 \hbar^2 k_y^2} \ ,
\end{align}
where $\mathbf{I}$ is the identity matrix.
Disorder is perturbatively incorporated in the averaged Green's
$\bar{G}$
function through a self-energy matrix $\Sigma(\vec{k},\epsilon)$ such that
\begin{equation}
\overline{G}^{R/A}(\vec{k},\epsilon)=\left[(\epsilon \mp i 0) \, \mathbf{I} -H^0(\vec{k})  \mp i ~ \textrm{Im}~ \Sigma (\vec{k},\epsilon)\right]^{-1}.
\end{equation}
The real part of the self-energy has been neglected. The elastic
scattering rates will be defined below from the imaginary part of the
self-energy.
  To lowest order in the disorder strength $\gamDes$, this self-energy, solution of a Dyson equation, reads
\begin{equation}
\Sigma (\vec{k} {, \epsilon}) = \int \frac{d\vec{k}'}{(2\pi)^2}
~\overline{ V(\vec{k}')V(-\vec{k}') }   ~ G^0(\vec{k}-\vec{k}'  {, \epsilon})  \ .
\end{equation}
Its imaginary part is then obtained as
\begin{align}
- \textrm{Im}~ \Sigma { (\epsilon)} &=
\pi \gamDes
\int_{-\theta_{0}}^{\theta_{0}}
d\theta ~ \rho(\epsilon,\theta) \left[ \mathbf{I} - \cos \theta ~ \sigma_x \right] \\
&= \frac{\hbar}{2 \tau_{e}^{0}{ (\epsilon)}} \left[ \mathbf{I} + r\left( \delta \right) \sigma_x \right] \ .
\label{eq:SigmaMatrix}
\end{align}
The densities of states $\rho(\epsilon,\theta)$ and $\rho(\epsilon)$, the bare scattering time $\tau_{e}^{0} (\epsilon)$
 and the anisotropy factor $r(\delta)$ have been defined in Eqs.~(\ref{eq:dos12},\ref{eq:defRho},\ref{eq:ElasticTimeMean},\ref{eq:defR}).
It is worth noting that this self-energy acquires an unusual matrix structure in  pseudo-spin space: this manifests within the diagrammatic
approach the anisotropy of the scattering time $\tau_e(\epsilon,\theta)$, which was described in Eq.~(\ref{eq:ElasticTimeParam})  previously.
  Indeed, in the Green function formalism, the direction of propagation of  eigenstates of the Hamiltonian (\ref{eq:PureModel}) is encoded into
  their spinor structure (the relative phase between their components, see Eq.~(\ref{eq:eigenstate})).
Hence the scattering time in the corresponding direction will be obtained as the matrix element of the above self-energy in the associated spinor eigenstate.

%%
%\begin{align}
%G^{R/A}(\vec{k},\epsilon) &= \left( \epsilon ~  \mathbf{I}  - H^0(\vec{k} - \pm \textrm{Im}~ \Sigma  ) \right)^{-1}
%\nonumber \\
%&=
%\frac{\epsilon ~  \mathbf{I} + \left( \frac{\hbar^2 k_x^2}{2m}+\Delta \right) \sigma^x  + c_y~ \hbar k_y \sigma^y }
%{(\epsilon \pm i \frac{\hbar}{2 \tau_e^0})^2  -\left( \frac{\hbar^2 k_x^2}{2m}+\Delta \mp i \frac{\hbar r(\Delta/\epsilon)}{2 \tau_e^0} \right)^2- c_y^2 \hbar^2 k_y^2}  .
%\end{align}

%
 %%----------------------------------  FIGURE  ---------------------------------------------%%
\begin{figure}
\begin{center}
\begin{tikzpicture}
[decoration=snake,
line around/.style={decoration={pre length=#1,post length=#1}},scale=1]
\draw[->,>=latex,thick] (0.,0) -- (1.5,0);
\node at (0.7,0.5) {$\left[ \overline{G}^{R} \right]_{ab}(\vec{k})$};
\node at (0.,-0.2) {$a$} ;
\node at (1.5,-0.2) {$b$} ;
\draw[<-,>=latex,dashed,thick] (2.5,0) -- (4.,0);
\node at (3.2,0.5) {$\left[ \overline{G}^{A} \right]_{cd}(\vec{k})$};
\node at (2.5,-0.2) {$d$} ;
\node at (4.,-0.2) {$c$} ;
\draw[thick,dotted] (5,0.) -- (7,0.);
\draw[fill=black!50] (6,0.) circle (0.15);
\node at (5.8,0.5) {$\overline{ V^{2}}(\vec{k})=\gamDes $};
\end{tikzpicture}
\end{center}
\caption{\label{fig:convention}
Conventions for the diagrammatic representation of perturbation theory of transport.
}
\end{figure}
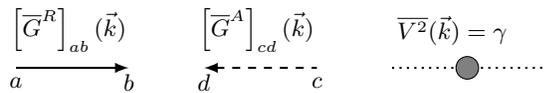
%%----------------------------------  FIGURE  ---------------------------------------------%%

%%----------------------------------  FIGURE  ---------------------------------------------%%
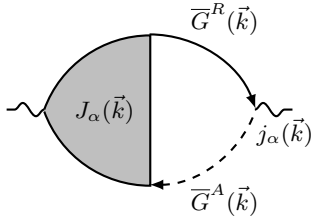
\begin{figure}[!ht]
\begin{center}
\begin{tikzpicture}
[decoration=snake,line around/.style={decoration={pre length=#1,post length=#1}},scale=1]
\begin{scope}
\draw[fill=black!50,fill opacity=0.5, thick]
	(0.5,0) arc (160:90:1.5)
	-- ++(0,-2)
	arc (270:200:1.5);
\end{scope}
\draw[-,decorate,  thick] (0.5,0) -- ++(-0.5,0);
\draw[-,decorate,  thick] (3.3,0) -- ++(+0.5,0);
\begin{scope}
\draw[->, >=latex, thick] (1.9,1) arc (90:20:1.5);
\draw[<-, >=latex,dashed,  thick] (1.9,-1) arc (270:340:1.5);
\end{scope}
\node at (1.3,0) {$J_\alpha(\vec{k})$};
\node at (2.9,1.2) {$\overline{G}^{R}(\vec{k})$};
\node at (2.9,-1.2) {$\overline{G}^{A}(\vec{k})$};
\node at (3.7,-.3) {$j_\alpha(\vec{k})$};
\end{tikzpicture}
\end{center}
\caption{\label{fig:classical conduct} Diagrammatic representation of the classical conductivity with the
conventions of Fig.~\ref{fig:convention}. The renormalized current operator
 is defined in Fig.~\ref{fig:renormCurrent}.
}
\end{figure}
%%----------------------------------  FIGURE  ---------------------------------------------%%

\subsection{Conductivity}
\label{sec:DiagConductivity}

\subsubsection{Kubo formula}

The longitudinal conductivity can be  deduced from the Kubo formula ($\alpha = x,y$) :
\begin{multline}
\sigma_{\alpha \alpha} = \\
\frac{\hbar}{2 \pi L^2} \mathrm{Tr} \left[
j_{\alpha}(\vec{k})
G^{R} (\vec{k},\vec{k}', \epsilon_{F})
j_{\alpha} (\vec{k}')
G^{A} (\vec{k}',\vec{k} , \epsilon_{F})
  \right],
 \label{eq:Kubo}
\end{multline}
where $\mathrm{Tr}$ corresponds to a trace over the pseudo-spin and momentum quantum numbers :
$\mathrm{Tr}=\mathrm{tr} ~\sum_{\vec{k}} \simeq L^2 \mathrm{tr}~  \int d\vec{k}/(2\pi)^2 $ and
$\mathrm{tr}$ is a trace over the pseudo-spin indices only.
For clarity, throughout this section on transport coefficients, we will omit the dependence on the Fermi energy $\epsilon_{F}$ of various quantities.
The current density operators are also operators acting on both spin and momentum spaces. They are
deduced from the Hamiltonian (\ref{eq:PureModel}) as:
\begin{equation}
j_x (\vec{k}) = -\frac{e}{m} \hbar k_x ~\sigma_x
\quad ; \quad
j_y (\vec{k})= -e c_y ~\sigma_y .
\end{equation}
Note that $j_x$ is linear in momentum while $j_y$ depends only on spin quantum numbers.
 Perturbation in the disorder amplitude of the conductivity ~(\ref{eq:Kubo}) is obtained by expanding the Green's function in the disorder potential $V$ before
 averaging over the gaussian distribution.
In the classical diffusive limit, the dominant terms which determine the averaged classical conductivity  are
 represented diagrammatically on Fig.~\ref{fig:classical conduct} and lead to
\begin{equation}
\overline{\sigma}_{\alpha \alpha} =
\frac{\hbar}{2 \pi L^2} \mathrm{Tr} \left[
J_{\alpha}
\overline{G}^{R}
j_{\alpha}
\overline{G}^{A}
  \right]
 \label{eq:ConductRenorm}
\end{equation}
where $J_\alpha$ is the renormalized current density operator. The discrepancy between $J_\alpha$ and the bare
current operator $j_\alpha$ accounts for the appearance of transport time
$\tau^{\textrm{tr}}_{\alpha}$ in the Boltzmann
approach\cite{McCann:2006} due to the anisotropy of scattering.
 This renormalized current operator is easier to define diagrammatically, as shown on Fig.~\ref{fig:renormCurrent}.

 %%----------------------------------  FIGURE  ---------------------------------------------%%
\begin{figure*}[!ht]
\begin{center}
\begin{tikzpicture}
[decoration=snake,line around/.style={decoration={pre length=#1,post length=#1}},scale=1]
\begin{scope}
\draw[-,decorate,  thick] (5.5,0) -- ++(-0.5,0);
\draw[fill=black!50,fill opacity=0.5, thick]
	(5.5,0) arc (160:90:1.5)
	-- ++(0,-2)
	arc (270:200:1.5);
\end{scope}
\node at (6.3,0) {$J_\alpha(\vec{k})$};
\node[scale=1] at (7.5,0.) {$=$};
\draw[-,decorate,  thick] (8.5,0) -- ++(-0.5,0);
\draw[-,thick] (8.5,0) arc (160:150:1.5);
\draw[-,thick] (8.5,0) arc (-160:-150:1.5);
\node[scale=1] at (9.,0.) {$+$};
\draw[-,decorate,  thick] (10.,0) -- ++(-0.5,0);
\draw[-,->,>=latex,thick] (10,0) arc (160:90:1.5);
\draw[-,<-,>=latex,dashed,thick] (10,0) arc (200:270:1.5);
\draw[-,dotted,thick] (11.4,1)--++(0,-2);
\draw[fill=black!50] (11.4,0.) circle (0.15);
\node[scale=1] at (12.,0.) {$+$};
\draw[-,decorate,  thick] (13.,0) -- ++(-0.5,0);
\draw[-,->,>=latex,thick] (13,0) arc (160:90:1.5);
\draw[-,<-,>=latex,dashed,thick] (13,0) arc (200:270:1.5);
\draw[-,->,>=latex,thick] (14.4,1)-- (15,1);
\draw[-,<-,>=latex,dashed,thick] (14.4,-1)-- (15,-1);
\draw[-,dotted,thick] (14.4,1)--++(0,-2);
\draw[-,dotted,thick] (15.,1)--++(0,-2);
\draw[fill=black!50] (14.4,0.) circle (0.15);
\draw[fill=black!50] (15.,0.) circle (0.15);
\node[scale=1] at (15.5,0.) {$+$};
\draw[-,decorate,  thick] (16.5,0) -- ++(-0.5,0);
\draw[-,->,>=latex,thick] (16.5,0) arc (160:90:1.5);
\draw[-,<-,>=latex,dashed,thick] (16.5,0) arc (200:270:1.5);
\draw[-,->,>=latex,thick] (17.9,1)-- (18.5,1);
\draw[-,->,>=latex,thick] (18.5,1)-- (19.1,1);
\draw[-,<-,>=latex,dashed,thick] (17.9,-1)-- (18.5,-1);
\draw[-,<-,>=latex,dashed,thick] (18.5,-1)-- (19.1,-1);
\draw[-,dotted,thick] (17.9,1)--++(0,-2);
\draw[-,dotted,thick] (18.5,1)--++(0,-2);
\draw[-,dotted,thick] (19.1,1)--++(0,-2);
\draw[fill=black!50] (17.9,0.) circle (0.15);
\draw[fill=black!50] (18.5,0.) circle (0.15);
\draw[fill=black!50] (19.1,0.) circle (0.15);
\node[scale=1] at (20.,0.) {$+ \cdots$};
\end{tikzpicture}
\begin{tikzpicture}
[decoration=snake,line around/.style={decoration={pre length=#1,post length=#1}},scale=1]
\begin{scope}
\draw[-,decorate,  thick] (5.5,0) -- ++(-0.5,0);
\draw[fill=black!50,fill opacity=0.5, thick]
	(5.5,0) arc (160:90:1.5)
	-- ++(0,-2)
	arc (270:200:1.5);
\end{scope}
\node at (6.3,0) {$J_\alpha(\vec{k})$};
\node at (7.1,1) {$a$};
\node at (7.1,-1) {$b$};
\node[scale=1] at (7.5,0.) {$=$};
\draw[-,decorate,  thick] (8.5,0) -- ++(-0.5,0);
\draw[-,thick] (8.5,0) arc (160:150:1.5);
\draw[-,thick] (8.5,0) arc (-160:-150:1.5);
\node at (8.5,-1) {$j_\alpha(\vec{k})$};
\node at (8.7,0.4) {$a$};
\node at (8.7,-0.4) {$b$};
\node[scale=1] at (9.,0.) {$+$};
\draw[-,decorate,  thick] (10.,0) -- ++(-0.5,0);
\draw[fill=black!50,fill opacity=0.5, thick]
	(10,0) arc (160:90:1.5)
	-- ++(0,-2)
	arc (270:200:1.5);
\draw[-,->,>=latex,thick] (11.4,1)-- (12.1,1);
\draw[-,<-,>=latex,dashed,thick] (11.4,-1)-- (12.1,-1);
\draw[-,dotted,thick] (12.1,1)--++(0,-2);
\draw[fill=black!50] (12.1,0.) circle (0.15);
\node at (10.8,0) {$J_\alpha(\vec{k}')$};
\node at (11.7,1.4) {$\overline{G}^R(\vec{k}')$};
\node at (11.7,-1.5) {$\overline{G}^A(\vec{k}')$};
\node[scale=1] at (12.7,0.) {$\gamDes$};
\node at (12.3,1) {$a$};
\node at (12.3,-1) {$b$};
\node at (11.6,0.85) {$c$};
\node at (11.6,-0.8) {$d$};
\end{tikzpicture}
\end{center}
\caption{\label{fig:renormCurrent}
 Schematic representation of renormalized current operator $[J_\alpha]_{ab}(\vec{k})$ as the infinite sum of vertex corrections to the
 bare current operator (top), and corresponding recursive equation satisfied by  $J_\alpha$ (bottom).
 }
\end{figure*}
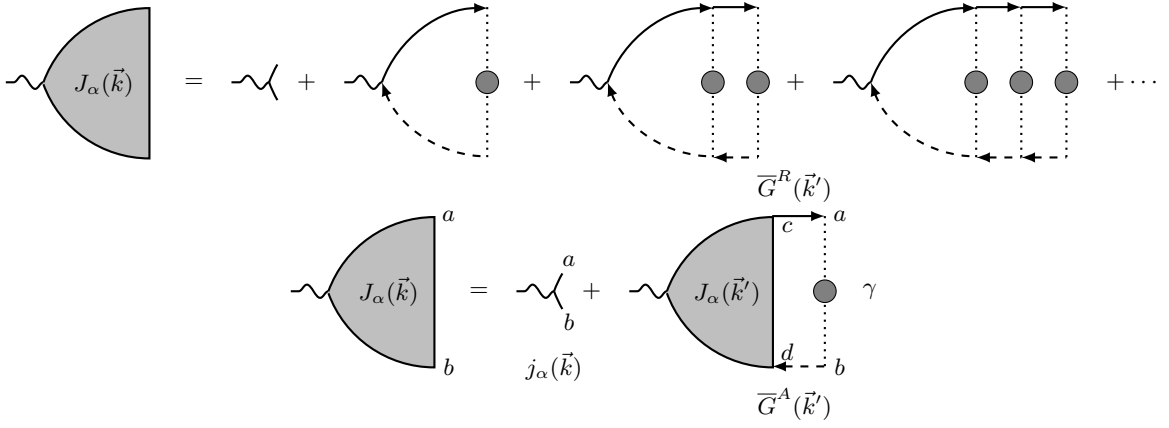
%%----------------------------------  FIGURE  -----------------------------------f----------%%
%

\subsubsection{Conductivity along $x$}

In this direction, the current operator is linear in $k_{x}$, while the averaged Green's functions
$\overline{G}^{R} (\vec{k}),\overline{G}^{A} (\vec{k})$ are even functions of $k_x$. Hence all the terms in the expression of
 the renormalized current $J_x$ with at least a Green's function vanish by $k_x \rightarrow -k_x$ symmetry, and
\begin{equation}
J_x(\vec{k}) = j_x(\vec{k}) =-\frac{e}{m} \hbar k_x ~\sigma_x \ .
\label{eq:RenormJx}
\end{equation}
There is no renormalization of the current operator, in agreement with the result $\tau^{\textrm{tr}}_x = \tau_e$ from
the Boltzmann equation approach.
In the $x$ direction, the expression (\ref{eq:ConductRenorm}) reduces to
\begin{multline}
\overline{\sigma}_{xx} =
\left( \frac{\hbar e}{m} \right)^2
\frac{\hbar}{2 \pi L^2}
 \mathrm{Tr} \left[
 %\int \frac{d\vec{k}}{(2 \pi)^2} ~
 k_x^2
\sigma_x \overline{G}^{R} (\vec{k})
\sigma_x \overline{G}^{A} (\vec{k})
  \right] .%;
 \label{eq:SxxTemp}
\end{multline}
%
%where $ \mathrm{tr}$ runs over the spin indices only.
%
Using $L^{-2}\sum_{\vec{k}} \simeq \int \rho(\epsilon,\theta) d\epsilon d\theta $
and
the parametrization defined in Eq. (\ref{eq:param}) of the contours of constant energy $\epsilon$ we perform the
integration over energy to obtain
\begin{multline*}
\overline{\sigma}_{xx} =
\frac{e^2 \tau^{0}_e \epsilon_F}{m }
\int_{-\theta_0}^{+\theta_0} d\theta ~
\frac{ \rho(\epsilon , \theta)~ (\cos \theta - \delta)}{1+r(\delta) \cos \theta}
\\
 \times \mathrm{tr} \biggl[
\sigma_x
 \left[ \mathbf{I} + \cos \theta \sigma_x + \sin \theta \sigma_y \right]
 \sigma_x
 \left[ \mathbf{I} + \cos \theta \sigma_x + \sin \theta \sigma_y \right]
 \biggr].
% \label{eq:SxxTemp2}
\end{multline*}
Performing the spin trace first, we obtain
\begin{multline}
\overline{\sigma}_{xx} =
4 \frac{e^2 \tau^{0}_e \epsilon_F}{m }
\int_{-\theta_0}^{+\theta_0} d\theta ~
\frac{ \rho(\epsilon , \theta)~ (\cos \theta - \delta)}{1+r(\delta) \cos \theta}
\cos^2 \theta \  .
\label{eq:SxxTemp3}
\end{multline}
By using eq.~(\ref{eq:DirectionalDensityRewriting}) for the directional density of states
we recover exactly the integral expression for the result  (\ref{eq:sigma_xx_boltzmann}) of Boltzmann approach:
\begin{align}
\overline{\sigma}_{xx} =
\frac{ e^2  \hbar}{ \pi \gamDes} ~
 \frac{2 \epsilon  }{m  } ~
 \frac{ \mathcal{I}_3(\Delta/\epsilon)}{  {I}_1(\Delta/\epsilon)} \ .
\label{eq:sigma_xx_diag}
\end{align}

\subsubsection{Renormalized current operator along $y$}

 In the $y$ direction, the  current operator is renormalized : the bare current operator $j_y$ is independent of the momentum
 $\vec{k}$ and the symmetry argument used for the $x$ direction does not hold anymore. This renormalized current operator satisfies a Bethe-Salpeter equation represented in Fig.~\ref{fig:renormCurrent}:
\begin{equation}
J_y  = j_y+ J_y  \Pi \gamDes
%\gamDes \left( \mathbf{I} \otimes \mathbf{I}  - \gamDes P_0  \right)^{-1},
 \label{eq:Bethe2}
\end{equation}
where tensor product in spin space are assumed and
\begin{equation}
\Pi(\epsilon, \Delta) =  \int \frac{d\vec{k}}{(2 \pi)^2}
\overline{G}^{R} (\vec{k},\epsilon)  \otimes \overline{G}^{A} (\vec{k},\epsilon )^T .
\end{equation}
Due to the spinorial structure of the wave functions,  this propagator is here an operator acting as the tensor product of two spin $\frac12$ spaces. The notation
$\cdots^T$ corresponds to a transposition of  spin matrices.
Using the parametrization defined in Eq. (\ref{eq:param}) of the contours of constant energy $\epsilon$ we perform the
integration over energy to obtain for $\Pi(\epsilon, \Delta) \equiv \Pi( \delta = \Delta / \epsilon)$:
\begin{multline}
\Pi(\delta) =
\frac{\pi \tau_e^0}{\hbar}
\int_{-\theta_0}^{+\theta_0} d\theta ~
\frac{ \rho(\epsilon , \theta)}{1+r(\delta) \cos \theta}
\\
 \times \left[ \mathbf{I} + \cos \theta \sigma_x + \sin \theta \sigma_y \right] \otimes
 \left[ \mathbf{I} + \cos \theta \sigma_x - \sin \theta \sigma_y \right] .
\label{eq:P0_3}
\end{multline}
The expression (\ref{eq:DirectionalDensityRewriting}) for the directional density of states allows to rewrite it as
\begin{multline}
\Pi(\delta) =
 \frac{1}{2 \gamDes I_1(\delta)} \biggl[
     \mathcal{I}_1(\delta) \mathbf{I} \otimes \mathbf{I}
 +  (\mathcal{I}_1(\delta) - \mathcal{I}_2(\delta))\sigma_x \otimes \sigma_x
\\
  - \mathcal{I}_2(\delta) \sigma_y \otimes \sigma_y
+\mathcal{J}_1(\delta)( \mathbf{I} \otimes \sigma_x + \sigma_x \otimes \mathbf{I}) \biggr] \ ,
\label{eq:P_D}
\end{multline}
where we introduced the functions:
\begin{align}
\mathcal{I}_1(\delta) &= \int_{-\theta_{0}}^{\theta_{0}} d \theta ~ \frac{1}{\sqrt{\cos \theta- \delta}\, (1+r(\delta) \cos \theta)}  \ , \\
%
%\mathcal{I}_6(x) = \int_{-\theta_{0}}^{\theta_{0}} d \theta ~ \frac{\cos^2 \theta}{\sqrt{\cos \theta-x}(1+ r(x) \cos \theta)} \\
%
\mathcal{J}_1(\delta) &= \int_{-\theta_{0}}^{\theta_{0}} d \theta ~ \frac{\cos \theta}{\sqrt{\cos \theta- \delta}\, (1+ r(\delta) \cos \theta)}\ ,
\end{align}
whereas $I_1$ and $\mathcal{I}_2$ are defined in Eqs.~(\ref{I1},\ref{eq:calI2}).

The inversion of the Bethe-Salpeter equation (\ref{eq:Bethe2}) is done in the appendix \ref{sec:appCurrent} and we find
\begin{equation}
J_y = j_y \left(\mathbf{I} \otimes \mathbf{I} - \gamDes \Pi(\delta)  \right)^{-1}
=  \left( 1 - \frac{\mathcal{I}_2(\delta) }{ I_1(\delta)}  \right)^{-1} j_y \ .
\label{eq:RenormJy}
\end{equation}

\subsubsection{Conductivity along $y$}

Following the formula (\ref{eq:ConductRenorm}), the average conductivity along $y$ is expressed
as
\begin{equation}
\overline{\sigma}_{yy} =
\frac{\hbar}{2 \pi } \mathrm{tr} \left[
J_{y}.\Pi(\delta) . j_y \right]  \ .
\label{eq:ConductRenormyy}
\end{equation}
From the eq.~(\ref{eq:Bethe2}), we express $J_y.\Pi(\delta)=\gamDes^{-1}(J_y-j_y)$ to obtain
from (\ref{eq:ConductRenormyy}):
\begin{equation}
\overline{\sigma}_{yy} =
\frac{\hbar}{2 \pi  \gamDes} \mathrm{tr} \left[(J_y-j_y) j_y \right] \ .
\end{equation}
The expression for the renormalized current operator (\ref{eq:RenormJy}) leads to the final result
\begin{equation}
\overline{\sigma}_{yy} =
\frac{e^2 \hbar}{ \pi  \gamDes} ~ c_y^2 ~
\frac{\mathcal{I}_2(\delta) }{ I_1(\delta) - \mathcal{I}_2(\delta)} \  ,
\label{eq:sigma_yy_diag}
\end{equation}
which is precisely the result (\ref{eq:sigma_yy_boltzmann}) obtained within the Boltzmann equation approach.

This concludes the derivation of the conductivity tensor within the diagrammatic approach. In doing so, we have identified the encoding of the anisotropic scattering
rates through the matrix self-energy (\ref{eq:SigmaMatrix}), while the corresponding transport times are hidden into the renormalization of vertex operators
(\ref{eq:RenormJx},\ref{eq:RenormJy}). Comparison with the Boltzmann approach allows to unveil the physical meaning of these technical structures, which we believe to be applicable to other situations of anisotropic transport of Dirac-like states.
 }%David%

\section{Results and Discussion}
\label{sect:discussion}

{
We now turn to a discussion of our results for various situations corresponding to energy spectra represented in Fig.~\ref{fig:EnergySpectrum}.
}

%Discussion from the expressions
%\begin{align}
%\sigma_{xx} &  =
% \frac{e^2 \hbar  }{\pi  \gamma} ~
% c_x^2  ~
% \left(
% \frac{\epsilon}{\Delta}~
%  \frac{ \mathcal{I}_3(\Delta/\epsilon)}{ \mathcal{I}_1(\Delta/\epsilon)} \right) ,
% \\
% \sigma_{yy} & =
% \frac{e^2 \hbar  }{\pi  \gamma} ~
% c_y^2 ~
% \frac{ \mathcal{I}_2(\Delta/\epsilon)}{ \mathcal{I}_1(\Delta/\epsilon) -  \mathcal{I}_2(\Delta/\epsilon) }
%\end{align}
%%
%where $c_x$ was defined after eq. ??? (model) and $\epsilon/|\Delta|$ is varied : we study the characteristics of different %phases.

 %Interpretation : dependance of $\sigma_{yy}$ on energy follows qualitatively the
 %energy dependence of $\lambda_y = \tau^{\textrm{tr}}/\tau_e$. Indeed, $c_y$ is independent of $\epsilon$
 %and $\rho(\epsilon) \tau_e(\epsilon)$ is approximately a constant of $\epsilon$.
 %Energy dependence follows from Einstein relation.

  %In the $x$ direction :  $\sigma_{xx}$ can be interpreted by the same Einstein relation but without any renormalization of
  %transport time with respect to elastic scattering time. The only energy dependance arises from the group velocity:
  %$v_x^2 \simeq \epsilon$.

\subsection{$\Delta =0$~: Semi-Dirac {spectrum} }

Focusing first on the merging point ($\Delta =0$), we find that the conductivities are expressed, from Eqs. (\ref{eq:sigma_xx_boltzmann},\ref{eq:sigma_yy_boltzmann},\ref{eq:sigma_xx_diag},\ref{eq:sigma_yy_diag}) as~:
\begin{subequations}
\begin{align}
\sigma_{xx} (\epsilon) &= \frac{e^2 \hbar  }{\pi \gamDes}\,  \frac{2  \epsilon}{m} \, \frac{  \mathcal{I}_3 (0)}{I_1(0)}  \simeq  0.197 \,   \frac{e^2 \hbar  }{\pi \gamDes}\,  \frac{2  \epsilon}{m}  \\
\sigma_{yy} (\epsilon) &= \frac{e^2 \hbar  }{\pi \gamDes}\,  c_y^2 \,  \frac{  \mathcal{I}_2 (0)}{I_1(0)- \mathcal{I}_2(0)}  \simeq  1.491 \frac{e^2 \hbar  }{\pi \gamDes}\,  c_y^2 \ .
\end{align}
\label{eq:sigma_SemiDirac}
\end{subequations}
This case, which corresponds to an hybrid dispersion relation, linear in one direction and quadratic in the other direction, has been previously studied in
Ref.~\onlinecite{Banerjee:12}. However these authors have neglected
both  the spinorial structure of the wave function and the angular
dependence of the elastic scattering time caused by the anisotropic
dispersion,
whose importance is emphasized in the present paper.
Using the numerical values of the integrals given in appendix B, we find $\lambda_y(0) \simeq 2.4915$ and\footnote{Note the correspondance between our notations
  $I_k(x), (k=1,2,3)$ and those $I_k$ of
  Ref.~\onlinecite{Banerjee:12} : $I_k(0)= 4 I_k$.}~:
\begin{subequations}
\begin{align}
\sigma_{xx} &= {\mathcal{I}_3(0) \over I_3(0)} ~ \sigma_{xx}^B \simeq 0.781 ~ \sigma_{xx}^B \\
\sigma_{yy} &= \lambda_y(0) {\mathcal{I}_2(0) \over I_2(0)}  ~ \sigma_{yy}^B  \simeq 2.237  ~ \sigma_{yy}^B \ ,
\end{align}
\end{subequations}
where $\sigma_{xx}^B$ and $\sigma_{yy}^B$ are the values obtained in Ref.~[\onlinecite{Banerjee:12}].

 The energy dependence of the conductivities (\ref{eq:sigma_SemiDirac})
 arises   from the energy dependence of the average squared velocities. It is therefore independent of the energy along the $y$ direction, while it
 increases linearly with energy along the $x$ direction.

\subsection{$\Delta >0$~: gapped {spectrum}}

When $\Delta >0$, the energy spectrum exhibits a gap and we study here the conductivity {above this gap} at energies $\epsilon > \Delta$.
Along the $x$ direction, the  renormalization factor $\lambda_x=1$ so that   $\tau_x^{tr}(\theta) = \tau_e (\theta)$. The energy dependence   arises mainly from the energy dependence of the average squared velocity.
 Therefore  we expect a roughly linear dependence in energy\footnote{More precisely, for $\ep \ll \Delta$ we have~:
$
\sigma_{xx} (\epsilon)
 \approx 0.197 \, e^2 \hbar  /(\pi \gamDes)  c_x^2 \,  (\epsilon/\Delta - 0.76)
$
}~:
\begin{eqnarray}
\sigma_{xx} (\epsilon) &=& \frac{e^2 \hbar  }{\pi \gamDes} \frac{2 \epsilon}{m}  \frac{  \mathcal{I}_3 (\Delta/\epsilon)}{I_1(\Delta/\epsilon)} = \frac{e^2 \hbar  }{\pi \gamDes} c_x^2 \, \frac{  \epsilon}{\Delta}  \frac{  \mathcal{I}_3 (\Delta/\epsilon)}{I_1(\Delta/\epsilon)} \\
&\approx&
0.2 \, \frac{e^2 \hbar  }{\pi \gamDes}  c_x^2 \, \frac{\epsilon - \Delta}{\Delta}
\label{eq:sigmaxx-highE-Delta-pos}
\end{eqnarray}
where
{
$c_x  = \sqrt{2  \Delta   / m}$ is the velocity along $x$ of the massive Dirac equation describing the spectrum for small momenta.
}%%%%

\begin{figure} [!h]
\centering
\includegraphics[width=8cm]{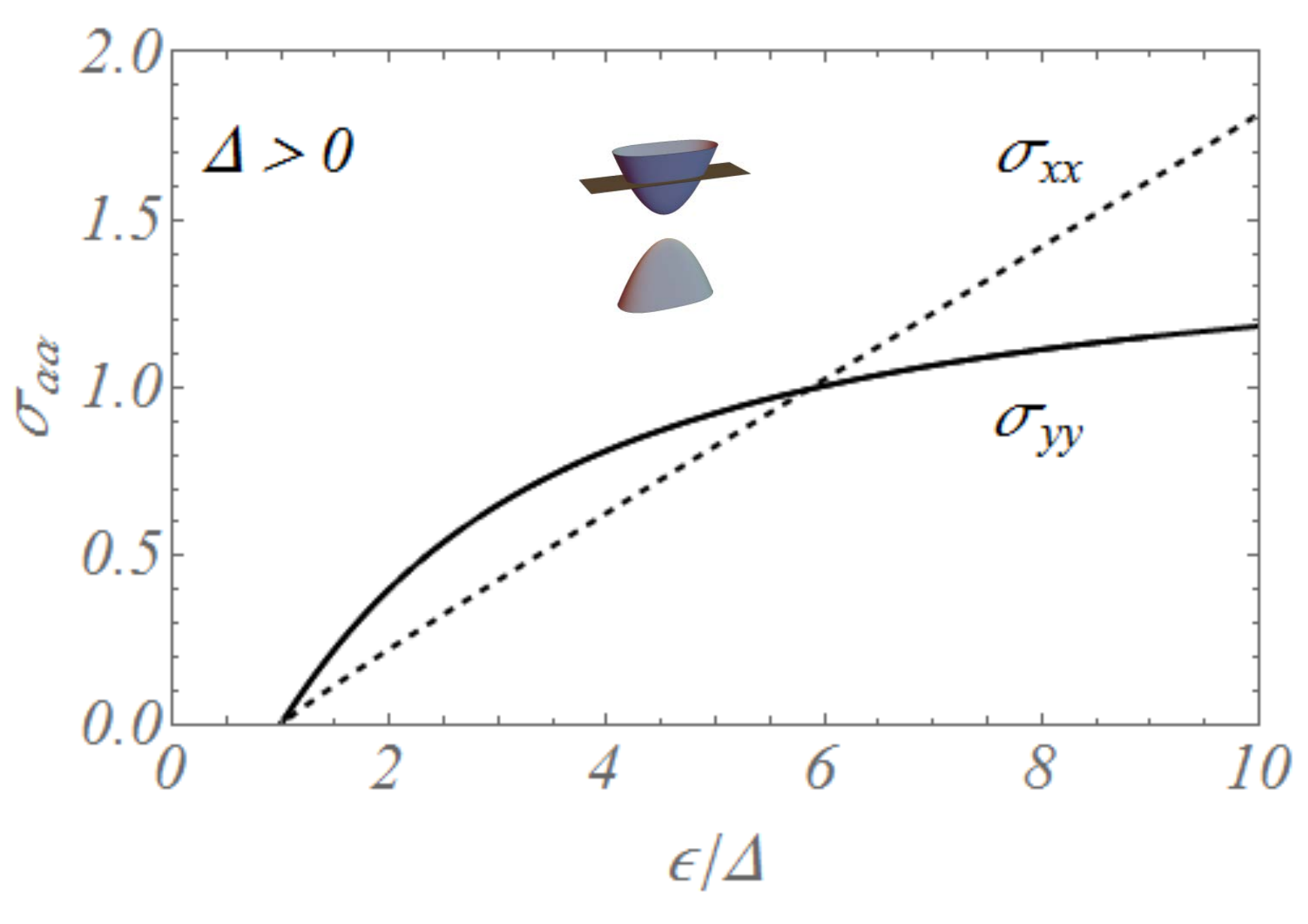}
\caption{(Conductivities $\sigma_{\alpha\alpha}$ in
units of $e^2 \hbar c_\alpha^2 / \gamDes$ for $\alpha = x,y$) as   functions of $\epsilon /\Delta $
 in the gapped phase ($\Delta > 0$).  The energy dependence of  $\sigma_{xx}$   arises from  the energy dependence of the velocity along $x$, while for $\sigma_{yy}$   it comes mainly from the energy dependence of the renormalization factor $\lambda_y$.
 }
\label{fig:sigmas_e_delta_positif}
\end{figure}

The dependence in energy of the conductivity $\sigma_{yy}$ is mainly due to the energy dependence of the renormalization factor $\lambda_y$ between transport time and relaxation time~:
\begin{eqnarray}
\sigma_{yy} (\epsilon) &=& \frac{e^2 \hbar }{\pi \gamDes} c_y^2  \frac{  \mathcal{I}_2 (\Delta/\epsilon)}{I_1(\Delta/\epsilon)- \mathcal{I}_2(\Delta/\epsilon)} \\
&\approx& \frac{e^2 \hbar  }{\pi \gamDes} c_y^2 \frac{  \mathcal{I}_2(0)}{I_1(0)}\,  \lambda_y({\Delta/\epsilon}) \ .
 \label{eq:sigmayy-highE-Delta-pos}
 \end{eqnarray}
The energy dependence of the conductivities  $\sigma_{xx}$ and $\sigma_{yy}$  is plotted in Fig.~\ref{fig:sigmas_e_delta_positif}.

%%%%%%%%%%%%%%%%%%%%%%%%%%%%%%%%%%%%%%%%%
%%%%%%%%%%%%%%%%%%%%%%%%%%%%%%%%%%%%%%%%%

\subsection{$\Delta <0$  : Dirac {spectrum} }

%%%%%%%%%%%%%%%%%%%%%%%%%%%%%%%%%%%%%%%%%
\begin{figure} [!h]
\centering
\includegraphics[width=8cm]{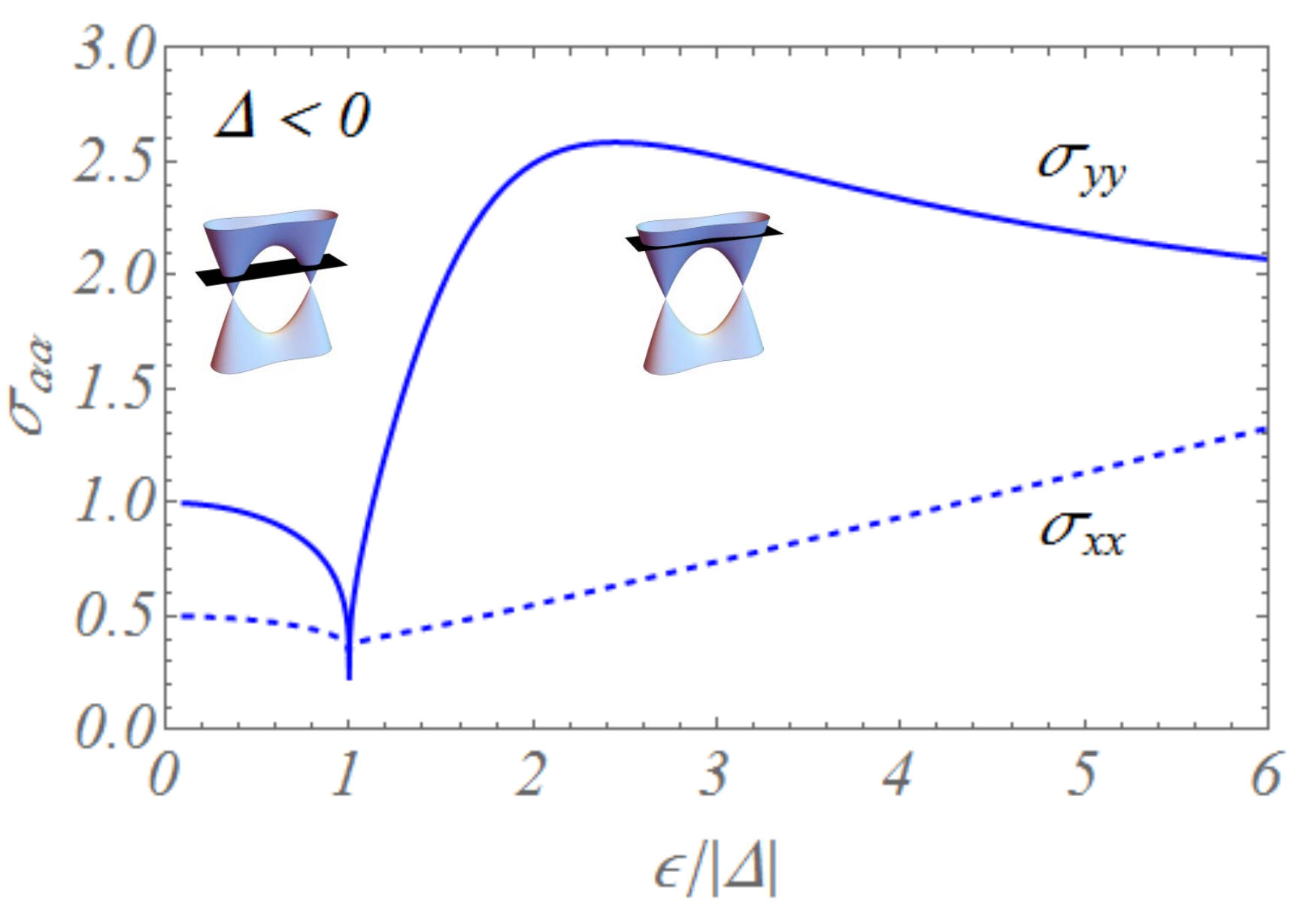}
\caption{Conductivities $\sigma_{\alpha\alpha}$ in
units of $e^2 \hbar c_\alpha^2 / \gamDes$ for $\alpha = x,y$) as   functions of $\epsilon /|\Delta |$
 in the Dirac phase
($\Delta <0$). The conductivity vanishes at the saddle point ($\epsilon=|\Delta|$). The vicinity of the saddle point should be treated with a self-consistent Born approximation (see text).
 }
\label{fig:sigmas_e_delta_negatif}
\end{figure}
%%%%%%%%%%%%%%%%%%%%%%%%%%%%%%%%%%%%%%%%%

In this phase, we have two regimes  separated  by the saddle point energy $|\Delta|$.
  At high energy above the saddle point, $\epsilon \gg |\Delta|$, the energy dependence of the conductivities are still given by Eqs.
 (\ref{eq:sigmaxx-highE-Delta-pos}, \ref{eq:sigmayy-highE-Delta-pos}).
In the low energy limit $\epsilon \ll \vert \Delta \vert$, expanding these expressions using Eq. (\ref{limits}), we recover the conductivities associated to a conic
dispersion  of characteristic velocities $c_x$ and $c_y$~:
\begin{eqnarray}
\sigma_{xx} (\epsilon \to 0) &=& \frac{e^2 \hbar  }{\pi  \gamDes} \, c_x^2\\
\sigma_{yy} (\epsilon \to 0) &=& 2 \frac{e^2 \hbar  }{\pi \gamDes} \, c_y^2  \ .
\end{eqnarray}
Note however the factor $2$ between the two expressions. This is due to the fact that $\tau^{\tr}_y= 2 \tau_e$ like in graphene
 while $\tau^{\tr}_x=  \tau_e$.  It is instructive to compare with this limit with the case of graphene, where it is known that
 $\tau^{\tr}= 2 \tau_e$ is all directions\cite{McCann:2006}, and where Einstein relation
 $\sigma_{\alpha\alpha}= e^2 (c_\alpha^2 \tau^{\tr} / 2) \rho(\ep_F)$ together with the Fermi golden rule
 $\tau^{\tr}= 2 \tau_e= 2 \hbar /(\pi \rho (\epsilon_F) \gamDes)$ leads to
 \begin{equation}
\sigma(\textrm{graphene}) = \frac{e^2 \hbar  }{\pi  \gamDes} \, v_F^2  \ . \\
\end{equation}
 Using the fact that $c_x^2 = c_y^2 = v_F^2/2$, we find the same result for $\sigma_{yy}$ but the conductivity is twice smaller
 along the $x$ direction.
   The difference by a factor 2 between $\sigma_{xx}$ and $\sigma_{yy}$
 results from intervalley scattering taking place along the $x$ direction (see Ref.~\onlinecite{McCann:2006} or a related discussion of diffusion within
 graphene with different intervalley and intravalley disorder rates).

It is important to note that our calculations predict vanishing
conductivities at the saddle point $\epsilon = \vert \Delta
\vert$. That is the result of the logarithmic divergence of the density
of states producing a vanishing elastic scattering time in
Eq.~(\ref{eq:ElasticTimeParam}). However, in such a limit, $ k_F l_e
\to 0$, so our approximations are no longer valid. To describe
correctly the behavior of the scattering time in the vicinity of the
saddle point, it is necessary to go beyond second order perturbation
theory, using for instance the self-consistent Born
approximation\cite{carpentier2013}, in order to obtain a finite
density of states and a non-zero elastic scattering
time. Qualitatively, we expect that the zero of the conductivity will
be replaced by a minimum for $\epsilon \simeq |\Delta|$.

   \subsection{Evolution of conductivites across the transition}
   \label{sec:transition}
\begin{figure}[h]
  \centering
  \includegraphics[width=9cm]{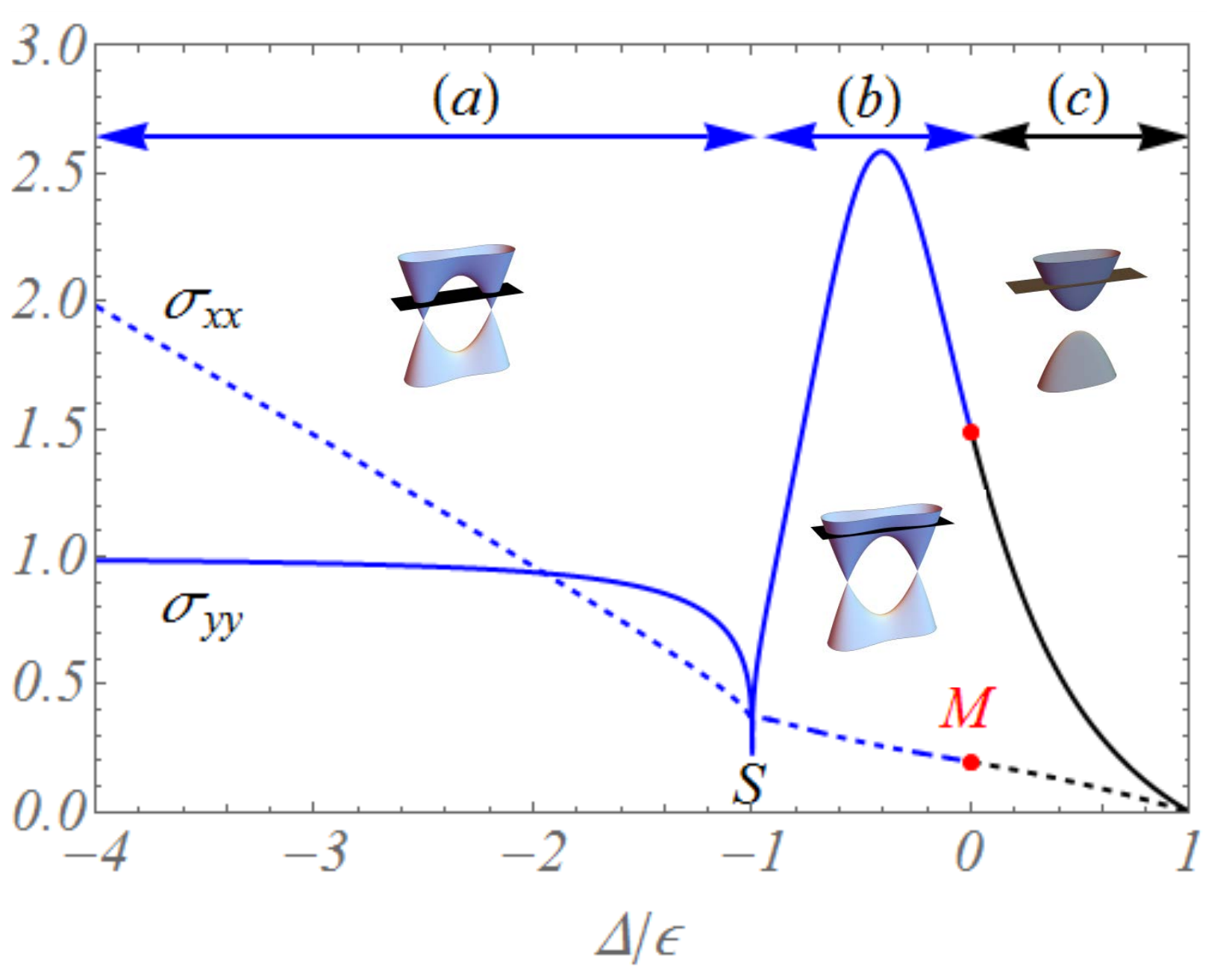}
  \caption{Conductivities  (in
units of $e^2 \hbar \tilde{c}_\alpha^2 / \gamDes$ for $\alpha = x,y$) as a function of the parameter $\Delta$ for
    a fixed chemical potential $\epsilon$. The conductivity
    $\sigma_{xx}$   decreases monotonically with $\Delta$,
    in almost linear fashion with a change of slope at the saddle
    point S. The conductivity $\sigma_{yy}$ is almost constant below
    the saddle point. Above the saddle point, the behavior of $\sigma_{yy}$ becomes non-monotonous with
  $\Delta$. At the metal-insulator transition, both $\sigma_{xx}$ and
  $\sigma_{yy}$ vanish linearly. Symbols (a), (b), (c), M and S refer to the regions presented in Fig.  \ref{fig:EnergySpectrum}.  }
  \label{fig:sigma-delta}
\end{figure}

We are now in position to discuss the evolution of the conductivity at fixed energy $\ep_F$, as
a function of the parameter $\Delta$ as we cross the merging
transition. Such evolution, derived for eq.~(\ref{eq:sigma_xx_boltzmann},\ref{eq:sigma_yy_boltzmann}) is represented on
Fig.~\ref{fig:sigma-delta}, where we have plotted $\sigma_{\alpha \alpha}$ in
units of $e^2 \hbar \tilde{c}_\alpha^2 / \gamDes$ for $\alpha = x,y$ and
$\tilde{c}_x  = \sqrt{2  \epsilon_F   / m}$ and $\tilde{c}_{y}=c_y$.   Below the saddle point for
$\Delta < -\epsilon_F$, $\sigma_{yy}$ is nearly constant, while
$\sigma_{xx}$ decreases almost linearly with $\Delta$.
At the saddle point $\Delta = -\epsilon_F$ where the topology of the Fermi surface changes,
 a dip in both $\sigma_{xx}$ and $\sigma_{yy}$ is visible, down to minimal values not quantitatively captured by the present approach.
 Past the saddle point, while $\sigma_{xx}$ remains linearly decreasing
with $\Delta$, albeit more slowly, $\sigma_{yy}$ is first increasing,
presenting a maximum for $\Delta/\epsilon_F \simeq -0.39$ and then
decreases to zero. No signature of the underlying transition at $\Delta =0$ is manifest in the transport at high Fermi energy $\epsilon_F$.

{
\section{Conclusion}

 We have studied the behavior of the conductivity in both the Dirac phase, the critical semi-Dirac and above the gapped phase.
 Using the complementary Boltzmann and diagrammatic techniques we have identified the different nature of anisotropy of the elastic scattering times
  and transport times.
   Indeed the transport is inherently anisotropic due both to the spinorial structure of the eigenstates and the anisotropy of the dispersion relation.
 The approaches developed in this paper can be generalized to study the diffusive transport in other semi-metallic phases, including the various three dimensional
 species recently identified.
}

\begin{acknowledgments}
We thank the hospitality of the Institut Henri Poincar{\'e} where part of this work was completed.
This work was supported by the French Agence Nationale de la Recherche (ANR) under grants SemiTopo (ANR-12- BS04-0007), IsoTop (ANR-10-BLAN-0419).
\end{acknowledgments}
\appendix

{

\section{Current vertex renormalization}
\label{sec:appCurrent}
\subsection{Inverse of tensor  product}

We use the notation $M^{ab}_{cd} = A_{ab} \otimes B_{cd} $ for the coefficients of a tensor product $M=A\otimes B$.
The inverse (for the outer product) $N=A^{-1}\otimes B^{-1}$ of $M$ satisfies the relation
$ M^{ab}_{cd} N^{be}_{df} = \delta_{ae} \delta_{cf} $.
In the obtention of the diffuson propagator, we need to invert a tensor product of the form
\begin{multline}
M =
     a~ \mathbf{I} \otimes \mathbf{I}
 +  b~\sigma_x \otimes \sigma_x
 + c~  \sigma_y \otimes \sigma_y
\\
+d~( \mathbf{I} \otimes \sigma_x + \sigma_x \otimes \mathbf{I}) \ .
\label{eq:KR_1}
\end{multline}
Its inverse $M^{-1}$ can be parametrized as
\begin{multline}
 M^{-1} = \Delta^{-1} \biggl[
     A~ \mathbf{I} \otimes \mathbf{I}
 +  B~\sigma_x \otimes \sigma_x
  + C~  \sigma_y \otimes \sigma_y
\\
+D~( \mathbf{I} \otimes \sigma_x + \sigma_x \otimes \mathbf{I})
+ E~ \sigma_z \otimes \sigma_z \biggr]  \ ,
\label{eq:P_2}
\end{multline}
with
\begin{subequations}
\begin{align}
A &=a^3 + 2 b d^2 - a (b^2 + c^2 + 2 d^2)   \\
B &= - b (a^2 - b^2 + c^2) + 2 (a - b) d^2   \\
C &=  c ~(-a^2 - b^2 + c^2 + 2 d^2)   \\
D &= - d \left[ (a - b)^2 - c^2 \right]    \\
E &= 2 c ~(-a b + d^2)   \\
\Delta &= \left[ (a - b)^2 - c^2 \right] \left[ (a + b)^2 - c^2 - 4 d^2 \right]  \ .
\end{align}
\end{subequations}
Let us now focus on the following contraction
%\begin{align}
%\left( \sigma_y\right)_{ab} \left[ M^{-1} \right]^{bc}_{ad} \left( \sigma_y \right)_{cd}
%&= 2~ \frac{A - B - C - E}{\Delta}  \\
%&= \frac{2}{a - b - c}.
%\label{eq:KronContraction}
%\end{align}
%
\begin{align}
\left[ M^{-1} \right]^{ab}_{cd}  \left( \sigma_y \right)_{bd}
&= \frac{A - B - C - E}{\Delta} \left( \sigma_y \right)_{ac}  \\
&= \frac{1}{a - b - c}  \left( \sigma_y \right)_{ac} \ .
\label{eq:KronContraction2}
\end{align}
irrespective of $d$ and hence of $\Delta$. In particular eq.~(\ref{eq:KronContraction2}) is  valid even if $\Delta$ vanishes.

\subsection{Current renormalization}

Let us now use the above parametrization (\ref{eq:KR_1}) for the tensor
$M=\mathbf{I} \otimes \mathbf{I} - \gamDes \Pi(\delta)$.
%where
%%
%\begin{multline}
%\Pi(\delta) =
% \frac{1}{2 \gamDes I_1(\delta)} \biggl[
%     \mathcal{I}_1(\delta) \mathbf{I} \otimes \mathbf{I}
% +  (\mathcal{I}_1(\delta) - \mathcal{I}_2(\delta))\sigma_x \otimes \sigma_x
%\\
%  - \mathcal{I}_2(\delta) \sigma_y \otimes \sigma_y
%+\mathcal{J}_1(\delta)( \mathbf{I} \otimes \sigma_x + \sigma_x \otimes \mathbf{I}) \biggr] \ ,
%\label{eq:P_D_Appendix}
%\end{multline}
%to obtain the contractions:
%\begin{align}
%\mathrm{Tr} \left( \sigma_y \Pi \sigma_y \right)  &=
%\left( \sigma_y\right)_{ab} \left[ \Pi \right]^{bc}_{ad} \left( \sigma_y \right)_{cd}  =
% \frac{ 2 \mathcal{I}_2(\delta)  }{\gamDes I_1(\delta)}
%\end{align}
%
%
%
%\subsection{Diffuson structure factor and contribution to the conductivity $\sigma_{yy}$.}
%
%Let us now use the above analysis to identify the structure factor for the Diffuson, defined in (\ref{eq:Bethe}) : we want to invert
%$M =  1 - \gamDes \Pi(\delta)$.
From the parametrization of eq.~(\ref{eq:P_D}), we obtain the following identification of coefficients
\begin{subequations}
\begin{align}
a &= 1 -  \frac{\mathcal{I}_1(\delta) }{2 I_1(\delta)} \  , &
b &= \frac{\mathcal{I}_2(\delta) - \mathcal{I}_1(\delta)  }{2 I_1(\delta)} \  , \\
c &= \frac{\mathcal{I}_2(\delta) }{2 I_1(\delta)} \ , &
d &= - \frac{\mathcal{J}_1(\delta) }{2 I_1(\delta)} \ .
\end{align}
\label{eq:CoefKron}
\end{subequations}
Then the equation (\ref{eq:KronContraction2})  provides the expression for the renormalized current operator:
\begin{equation}
J_y = j_y \left(\mathbf{I} \otimes \mathbf{I} - \gamDes \Pi(\delta)  \right)^{-1}
=  \left( 1 - \frac{\mathcal{I}_2(\delta) }{ I_1(\delta)}  \right)^{-1} j_y  \  .
\end{equation}
A word of caution is necessary at this stage :  $ \left( 1 - \gamDes \Pi(\vec{q}) \right)^{-1}$
is the structure factor which encodes the propagation of the diffuson modes\cite{Akkermans}.
In the symplectic class which we consider,  there is one such mode which is diffusive :  $1 - \gamDes \Pi(\vec{q})$
possesses a vanishing eigenvalue $\propto (Dq^2)$. Hence in the limit $q\to 0$ that we consider, $ 1 - \gamDes \Pi$ is no longer invertible.  In principle, we should have kept a finite momentum $q$ during the calculation, and
taken the limit $q\to 0$ only in the result.
However, the vertex renormalization that we consider in this section
is not sensitive to this long-wavelength physics : it corresponds to a renormalization of the elastic scattering time into a transport time, which occurs
on short distances. This is manifest in the independence of the result (\ref{eq:KronContraction2}) on the determinant $\Delta$ of the matrix $M$: this is a classical contribution, which depends on these massive diffuson modes, while the diffusive long
distance modes enter the quantum correction not discussed in this paper.

}%%blue%%

\section{Special Functions}

\begin{figure} [!h]
\centering
\includegraphics[width=4cm]{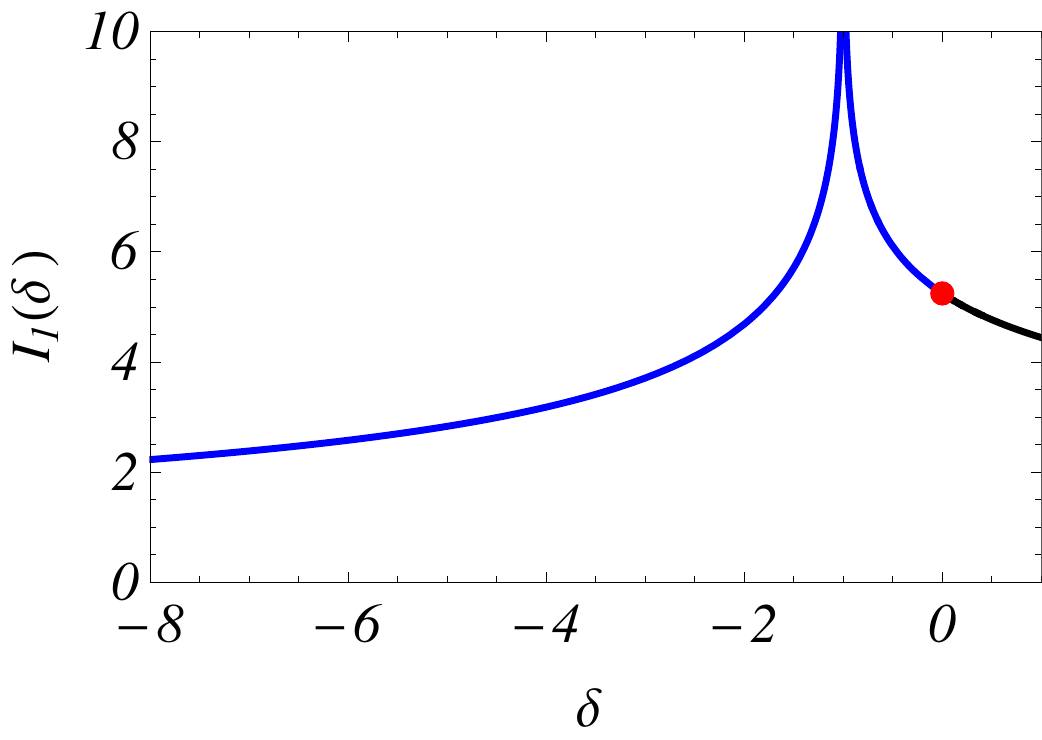}
\includegraphics[width=4cm]{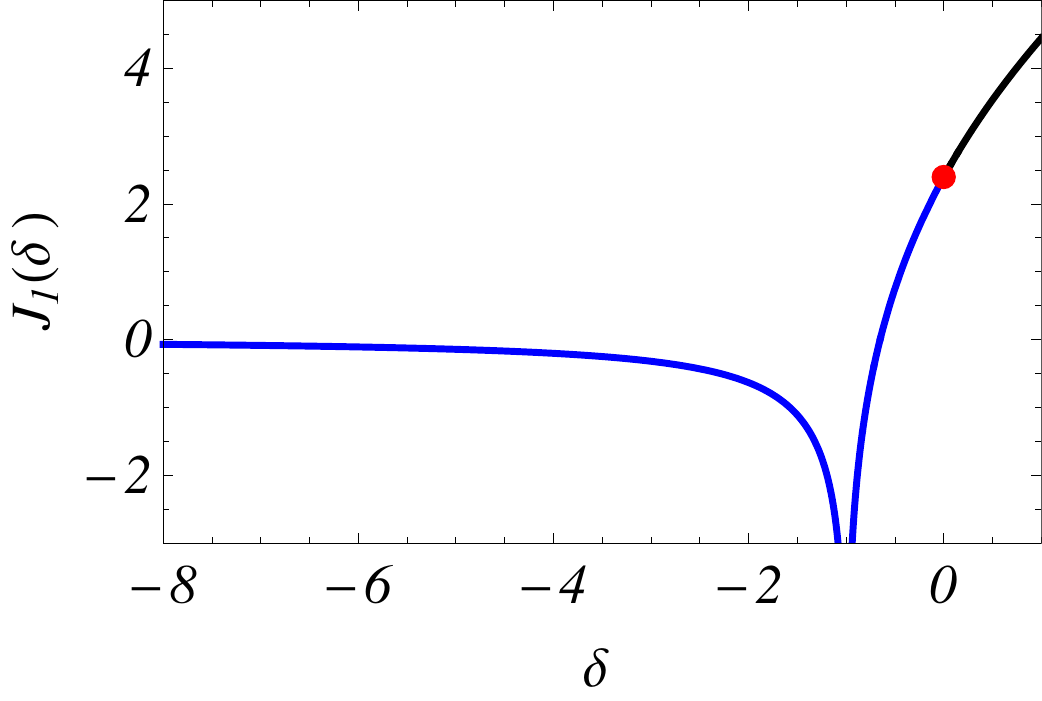}
\caption{   Functions ${I}_1(\delta)$ and ${J}_1(\delta)$ }
\label{fig:I1J1}
\end{figure}

In this appendix, we discuss a few useful results of the various integrals entering the expressions of transport coefficients in the paper and arising as
integrals along the constant energy contours of the model.
Let us first consider the integrals~:
\begin{equation*}
 {I}_1 (\delta)= \int_{-\theta_0}^{\theta_0}  {d \theta \over \sqrt{\cos \theta - \delta} }
 \quad , \quad
  {J}_1 (\delta)=  \int_{-\theta_0}^{\theta_0}  {d \theta \cos \theta
    \over \sqrt{\cos \theta - \delta} },
 \end{equation*}
with $\cos \theta_0=\delta$ if $|\delta|<1$ and $\theta_0=\pi$
otherwise.
 Defining $X= \sqrt{2 / (1 - \delta)}$, we find\footnote{We use the definition of the elliptic integrals  from  Gradshteyn and  Ryzhik \cite{Gradstein}. They differ from those used in Mathematica :
 $K_{Grad.}(x)= K_{Math.}(x^2)$,  $E_{Grad.}(x)= E_{Math.}(x^2)$ and  $\Pi_{Grad.}(\phi,n,x)=\Pi_{Math.}(n,\phi,x^2)$.}:
 \begin{eqnarray}
  {I}_1 (\delta)&=& {4 \over \sqrt{1 - \delta}} K(X)   \qquad  \delta<-1 \\
   {I}_1 (\delta)&=&2 \sqrt{2} K(1/X)  \qquad  -1 < \delta < 1
   \end{eqnarray}
and
\begin{multline}
  {J}_1 (\delta)= \\
  {4 \over \sqrt{1 - \delta}}[(1-\delta)  E(X)+ \delta K(X)]    \textrm{ for }  \delta<-1 \ ,
\end{multline}
and
\begin{multline}
   {J}_1 (\delta)= \\
   2 \sqrt{2}[2 E(1/X) - K(1/X)]   \textrm{ for }  -1 < \delta < 1 \  .
\end{multline}
The anisotropy factor $r(\delta)= J_1(\delta)/I_1(\delta)$ reads~:
\begin{align}
  r(\delta)&= (1 - \delta) E(X)/K(X) + \delta    &&  \textrm{ for }  \delta <-1 \\
  r(\delta)&= 2   E(1/X) / K(1/X) - 1  &&   \textrm{ for } -1 < \delta  < 1  \ .
\end{align}
The functions $I_1(\delta$ and $J_1(\delta)$ are plotted in Fig.~\ref{fig:I1J1}, and the dependence on $\delta$
of $r(\delta)$ is shown on Fig.~\ref{fig:r}.

%%%%%%%%%%%%%%%%%%%%%%%%%%%%%%%%%%%%%%%%
%%%%%%%%%%%%%%%%%%%%%%%%%%%%%%%%%%%%%%%%%
{

%%%%%%%%%%%%%%%%%%%%%%%%%%%%%%%%%%%%%%%%
\begin{figure} [!h]
\centering
\includegraphics[width=4cm]{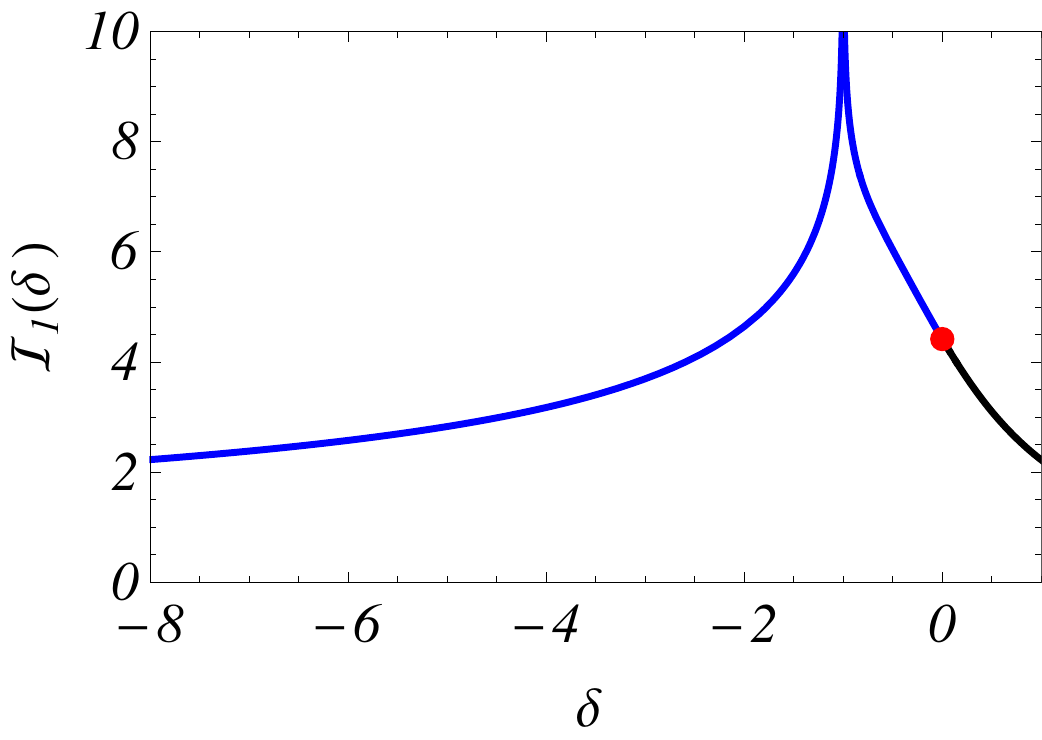}
\includegraphics[width=4cm]{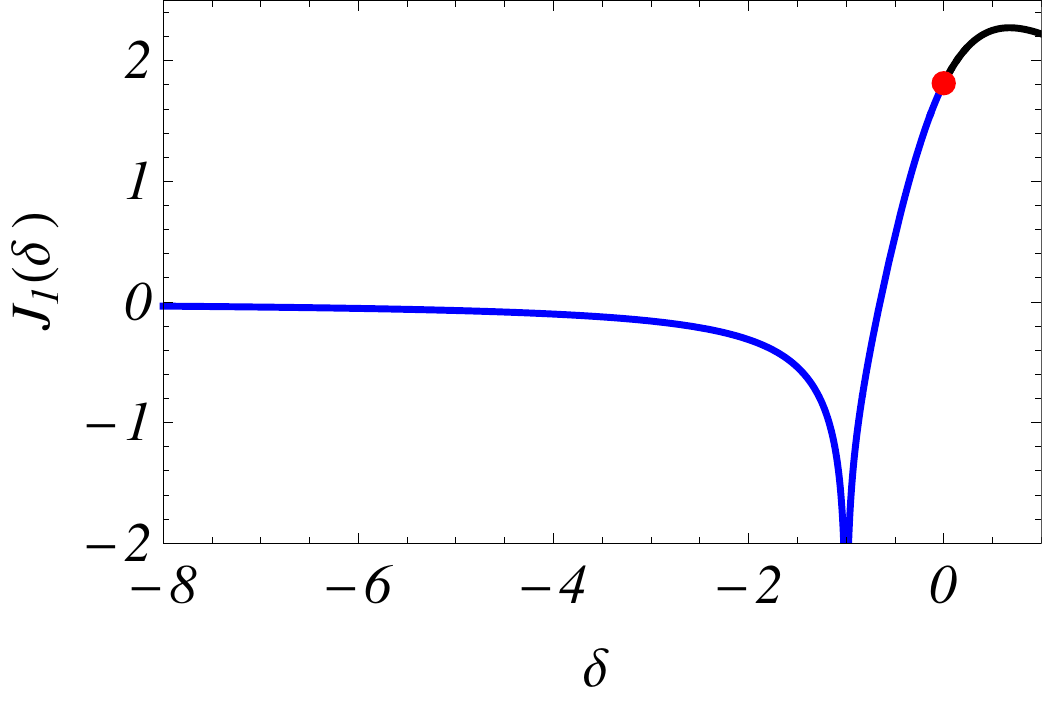}
\includegraphics[width=4cm]{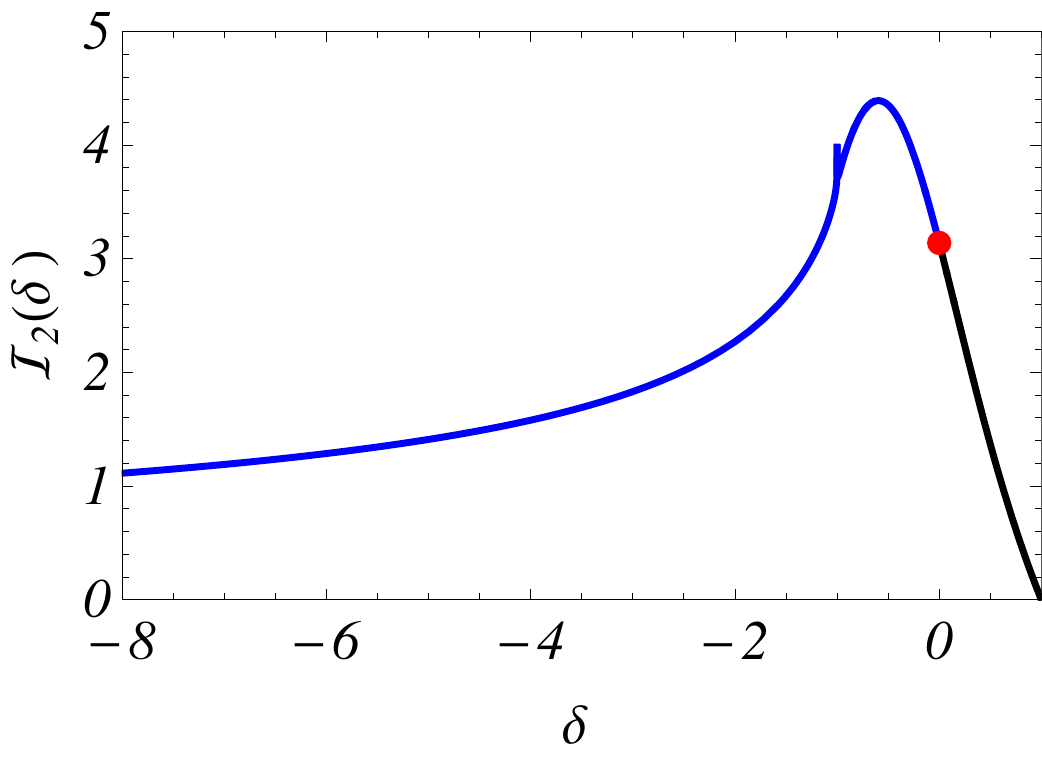}
\includegraphics[width=4cm]{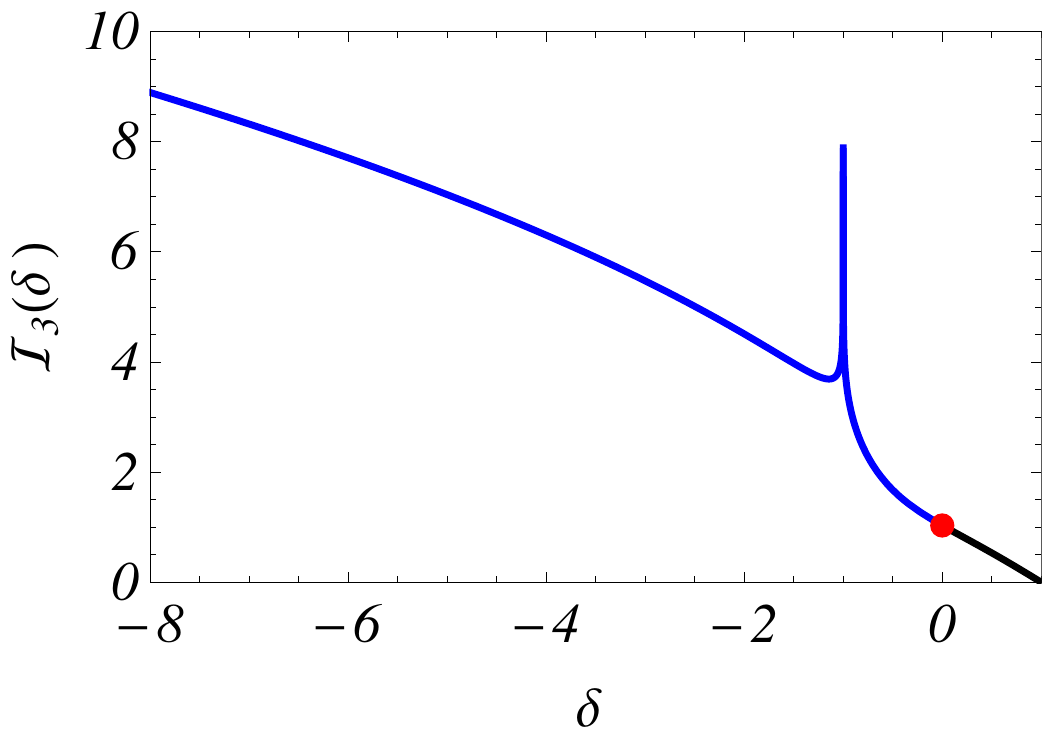}
\caption{ {Functions ${\cal I}_1(\delta)$, ${\cal J}_1(\delta)$, ${\cal I}_2(\delta)$ and ${\cal I}_3(\delta)$ }}
\label{fig:calIJ}
\end{figure}
%%%%%%%%%%%%%%%%%%%%%%%%%%%%%%%%%%%%%%%%%

Let us now consider the four integrals
\begin{align}
 \mathcal{I}_1(r,\delta) & =\int_{-\theta_0}^{\theta_0}  \frac{d\theta}{(1+r \cos \theta) \sqrt{\cos \theta-\delta}}
\label{eq:calligraphic-I1-integral}
\\
\mathcal{J}_1(r,\delta) & =\int_{-\theta_0}^{\theta_0} \frac{d\theta \cos \theta}{(1+r \cos \theta) \sqrt{\cos \theta-\delta}}
\label{eq:calligraphic-J1-integral}
\\
\mathcal{I}_2(r,\delta) &=\int_{-\theta_0}^{\theta_0} \frac{d\theta \sin^2 \theta}{(1+r \cos \theta) \sqrt{\cos \theta-\delta}}  \label{eq:calligraphic-I2-integral}
\\
\mathcal{I}_3(r,\delta)&=\int_{-\theta_0}^{\theta_0}\frac{d\theta \cos^2 \theta \sqrt{\cos \theta-\delta} }{(1+r \cos \theta) }
\label{eq:calligraphic-I3-integral}
\end{align}
}%%G%%
We first focus on first the integral $\mathcal{I}_1(r,\delta)$. For $\delta <-1$, it can be rewritten as:
\begin{align}
  \mathcal{I}_1(r,\delta)=&\frac{4}{(1+r) \sqrt{1-\delta}} \nonumber \\
  &
  \int_{0}^{\frac \pi 2}
  \frac{d\theta}{\left( 1-\frac{2r}{1+r} \sin^2 \theta\right)
    \sqrt{1-\frac{2}{1-\delta}\sin^2 \theta}} \\
   =& \frac{4}{(1+r) \sqrt{1-\delta}} ~\Pi\left(\frac \pi 2, \frac{2r}{1+r}, \sqrt{\frac 2
       {1-\delta}}\right)\ ,
\end{align}
where $\Pi$ is an elliptic integral of the third kind.
For $|\delta|<1$, we use the change of variable $\sin \frac \theta 2 =
\sin \frac{\theta_0} 2 \sin \varphi$ to rewrite
this function as:
\begin{align}
  \mathcal{I}_1(r,\delta)=&\frac{2 \sqrt{2}}{1+r} \nonumber \\
  \int_0^{\frac \pi 2} &
  \frac{d\varphi}{\left(1-\frac{2r}{1+r} \sin^2 \frac{\theta_0} 2
      \sin^2 \varphi\right) \sqrt{1-\sin^2 \frac{\theta_0} 2
      \sin^2 \varphi}} \nonumber \\
   =& \frac{2 \sqrt{2}}{1+r} ~\Pi\left(\frac \pi 2,
     \frac{2r}{1+r}\sin^2 \frac{\theta_0} 2,
     \sin \frac{\theta_0} 2 \right)\ .
\end{align}
Moreover the integrals $\mathcal{J}_1(r,\delta)$, $\mathcal{I}_2(r,\delta)$ and
$\mathcal{I}_3(r,\delta)$ can be expressed in terms of
$I_1(\delta)$, $J_1(\delta)$ and $\mathcal{I}_1(r,\delta)$:
{
\begin{align}
\mathcal{J}_1(r,\delta) &= \frac 1 r \left[ I_1(\delta) - \mathcal{I}_1(r,\delta)\right]\ , \\
 \mathcal{I}_2(r,\delta) &=   \frac 1 {r^2} I_1(\delta) - \frac 1 r  J_1(\delta)+ \left( 1-\frac 1 {r^2}\right)
  					\mathcal{I}_1(r,\delta)\ ,  \\
\mathcal{I}_3(r,\delta) &= \frac 1 r \left(\frac 1 3 + \frac \delta r
		+ \frac 1 {r^2}\right) I_1(\delta) - \frac 1 r \left(\frac{\delta}{3}
		+\frac 1 r\right) J_1(\delta)
 \nonumber \\
 & - \frac 1 {r^2} \left(\delta + \frac 1 {r}\right) \mathcal{I}_1(r,\delta) \ .
\end{align}
Finally, the integrals used in the text are
\begin{align*}
\mathcal{I}_1( \delta)&=\mathcal{I}_1[r(\delta),\delta]
&&
\mathcal{J}_1( \delta)=\mathcal{J}_1[r(\delta),\delta]   \\
\mathcal{I}_2( \delta)&=\mathcal{I}_2[r(\delta),\delta]
&&\mathcal{I}_3( \delta)=\mathcal{I}_3[r(\delta),\delta]  .
\end{align*}
The following special values are of particular interest for the expressions in the text :
\begin{align*}
I_1(0) & = 2 \sqrt{2} K\left(\frac{1}{\sqrt{2}}\right) \simeq 5.2441   \\
J_1(0)  & = \pi \sqrt{2} /  K\left(\frac{1}{\sqrt{2}}\right) \simeq 2.3963   \\
r(0) & =  J_1(0) /  I_1(0) \simeq 0.457   \\
I_2(0) & = {4 \over 3}  \sqrt{2} K(1/\sqrt{2}) \simeq 3.4961   \\
\mathcal{I}_2(0) & \simeq  3.1393   \\
I_3(0) & \simeq 1.4377   \\
\mathcal{I}_3(0) & \simeq 1.0322  \ .
\end{align*}
 as well as the limits when $\delta \rightarrow - \infty$~:
\begin{align}
\frac{  \mathcal{I}_2 (\delta)}{I_1(\delta)- \mathcal{I}_2(\delta)} &\to  1
;&&
\frac{  \mathcal{I}_3 (\delta)}{I_1(\delta)} \to  -  \frac{\delta}{2} \ .
\label{limits}
\end{align}

}%%G%%

%\bibliographystyle{apsrev4-1}
%\bibliography{SemiDirac}


 \end{document}